\documentclass{aa}

\pdfoutput=1

\usepackage{natbib}
\usepackage{txfonts}
\usepackage{gensymb}
\usepackage{inputenc}

\bibpunct{(}{)}{;}{a}{}{,}

\usepackage{color}
\usepackage{url}

\usepackage{caption}
\usepackage{subcaption}

\usepackage[breaklinks=true,colorlinks,citecolor=blue]{hyperref}

\usepackage{breakurl}

\newcommand{\ttt}{\texttt}
\newcommand{\mrm}{\mathrm}
\newcommand{\mtt}{\mathtt}
\newcommand{\WISC}{WISE $\times$ SCOS}
\newcommand{\WISCS}{WISE $\times$ SCOS $\times$ SDSS }

\newcommand{\tcb}{\textcolor{blue}}

\usepackage[normalem]{ulem}

\begin{document} 

   \title{Machine-learning identification of galaxies\\in the WISE $\times$ SuperCOSMOS all-sky catalogue}
  
  \titlerunning{Machine-learning identification of galaxies in \WISC} 
  
  \authorrunning{Krakowski et al.}

   \author{T. Krakowski\inst{1}, K. Ma{\l}ek\inst{1,2}, M. Bilicki\inst{3,2},  A. Pollo\inst{1,4,2},  A. Kurcz\inst{4,2}, M. Krupa\inst{4,2}}

   \institute{{National Centre for Nuclear Research, ul. Andrzeja Sołtana 7,  05-400 Otwock, Poland} \email{tomasz.krakowski@ncbj.gov.pl} 
   \and{Janusz Gil Institute of Astronomy, University of Zielona G\'ora, ul. Lubuska 2, 65-265 Zielona G\'ora, Poland} 
   \and{Leiden Observatory, Leiden University, P.O. Box 9513 NL-2300 RA  Leiden, The Netherlands}    
    \and{Astronomical Observatory of the Jagiellonian University, ul.\ Orla 171, 30-244 Krak\'{o}w, Poland }      
      }

   \date{\today}

  \abstract
{The two currently largest all-sky photometric datasets, WISE and SuperCOSMOS, have been recently cross-matched to construct a novel photometric redshift catalogue on 70\% of the sky. Galaxies were separated  from stars and quasars through colour cuts, which may leave imperfections because different source types may overlap in colour space.}
{The aim of the present work is to identify galaxies in the WISE $\times$ SuperCOSMOS catalogue through an alternative approach of machine learning. This allows us to define more complex separations in the multi-colour space than is possible with simple colour cuts, and should provide a more reliable source classification.}
 {For the automatised classification we used the support vector machines (SVM) learning algorithm and employed SDSS spectroscopic sources that we cross-matched with WISE $\times$ SuperCOSMOS to construct the training and verification set. We performed a number of tests to examine the behaviour of the classifier (completeness, purity, and accuracy) as a function of source apparent magnitude and Galactic latitude. We then applied the classifier to the full-sky data and analysed the resulting catalogue of candidate galaxies.   We also compared the resulting dataset with the one obtained through colour cuts.}
 {The tests indicate very high accuracy, completeness, and purity ($>95\%$) of the classifier at the bright end; this deteriorates for the faintest sources, but still retains acceptable levels of $\sim85\%$. No significant variation in the classification quality with Galactic latitude is observed.
When we applied the classifier to all-sky  WISE $\times$ SuperCOSMOS data, we found 15 million galaxies after
masking problematic areas. The resulting sample is purer than the one produced by applying colour cuts, at the price of a lower completeness across the sky.}
  {The automatic classification is a successful alternative approach to colour cuts for defining a reliable galaxy sample. The identifications we obtained are included in the public release of the WISE $\times$ SuperCOSMOS galaxy catalogue.\thanks{Available from \url{http://ssa.roe.ac.uk/WISExSCOS}.}}


   \maketitle
%

\section{Introduction }

Modern wide-field astronomical surveys include millions of sources, and future catalogues
will increase these numbers to billions. 
As most of the detected objects  cannot be followed-up spectroscopically, research done with such datasets will heavily rely on photometric information. 
Without spectroscopy, an appropriate identification of various source types is complicated, however. 
In the seemingly most trivial case of star-galaxy separation in deep-imaging catalogues, we quickly reach the limit where this cannot be done based on morphology: we lack resolution, and distant faint galaxies become unresolved or point-like, similar
to stars \citep[e.g.][]{vasconcellos11}. 
Additional information is then needed to separate out these and sometimes other  classes of sources (such as point-like but extragalactic quasars). 
This has traditionally been done with magnitude and colour cuts; however, when the parameter space is multidimensional, such cuts become very complex. 
Additionally, noise in photometry scatters sources from their true positions in the colour space. 
This, together with huge numbers of sources, most of which are usually close to the survey detection limit, precludes reliable overall identification with any manual or by-eye methods.  
For these reasons, the idea of automatized source classification has recently gained popularity and was applied to multi-wavelength datasets such as AKARI \citep{solarz12}, the Panoramic Survey Telescope and Rapid Response System \citep[Pan-STARRS,][]{saglia12}, the VIMOS Public Extragalactic Redshift Survey  \citep[VIPERS,][]{malek13SVM}, the cross-match of the Wide-field Infrared Survey Explorer  --  Two Micron All Sky Survey (WISE--2MASS)   \citep{KoSz15}, the Sloan Digital Sky Survey \citep[SDSS,][]{Brescia15}, and the WISE-only \citep{Kurcz16}, and it has been tested in view of the Dark Energy Survey data \citep{soumagnac13}. 

The present paper describes an application of a machine-learning algorithm to identify galaxies in a newly compiled dataset, based on the two currently largest all-sky photometric catalogues: WISE in the mid-infrared, and SuperCOSMOS in the optical. 
This work is a refinement of a simpler approach at source classification that was applied in \cite{WISC16}, hereafter \tcb{B16}, where stars and quasars were filtered out on a statistical basis using colour cuts to obtain a clean galaxy sample for the purpose of calculating photometric redshifts. 
The two parent catalogues we use here, described in detail below and in Sect.\ \ref{Sec: Data}, both include about a billion detections each, of which a large part are in common. 
For various reasons, however, the available data products from these two surveys offer limited information on the nature of the catalogued objects, which indeed presents a challenge to the classification task. 

WISE \citep{WISE}, which is the more sensitive of the two, suffers from low native angular resolution resulting from the small aperture of the telescope (40 cm): it is equal to $6.1''$ in its shortest $W1$ band (3.4 $\mu$m), increasing to $12''$ at the longest wavelength $W4$ of 23 $\mu$m. 
This leads to severe blending in crowded fields, such as at low Galactic latitudes, and original photometric properties of the blended sources become mixed. 
In addition, proper isophotal photometry has not been performed for the majority of WISE detections, and no WISE all-sky extended source catalogue is available as yet (see, however, \citealt{Cluver14} and \citealt{Jarrett16} for descriptions of ongoing efforts to improve on this situation).
Finally, WISE-based colours provide limited information for classification purposes: at rest frame, the light in its two most sensitive passbands, $W1$ and $W2$ (3.4 and 4.6 $\mu$m), is emitted from the photospheres of evolved stars (Rayleigh-Jeans tail of the spectrum), and the catalogue is dominated by stars and galaxies
of relatively low redshift, which typically have similar $W1-W2$ colours. 
The two other WISE filters, $W3$ and $W4$ centred on 12 and 23 $\mu$m, respectively which might serve to reliably separate out stars from galaxies and QSOs when combined with $W1$ and $W2$ \citep{WISE}, offer far too low detection rates to be applicable for most of the WISE sources.

SuperCOSMOS (hereafter SCOS) on the other hand, which is based on the scans of twentieth century photographic plates \citep{SCOS1}, does offer point and resolved source identification \citep{SCOS2}. 
This classification, although quite sophisticated, is based mostly on morphological information, however, and on the one hand, unresolved galaxies and quasars are classified as point sources, and on the other, blending in crowded fields (Galactic Plane and Bulge, Magellanic Clouds) leads to spurious extended source identifications (see also \citealt{Peacock16}). 

A cross-match of the WISE and SCOS catalogues improves the classification of different types of sources that is useful for extragalactic applications, as shown in \tcb{B16}. 
However, although only extended SCOS sources were considered in B16, blends mimicking resolved objects dominated at Galactic latitudes as high as $\pm30\degree$ and had to be removed on a statistical basis. 
In the present paper we improve on that work by generating a wide-angle (almost full-sky) galaxy catalogue from the \WISC\ cross-match through machine-learning. For this purpose we use the support vector machines (SVM) supervised algorithm.

A similar task for other WISE-based datasets was undertaken in two recent works. \cite{KoSz15}, who used a cross-match of WISE $W1<15.2$ sources with the 2MASS Point Source Catalogue \citep[PSC,][]{2MASS} and performed an SVM analysis in multicolour space, showed that a cut in the $W1_\mrm{WISE}-J_\mrm{2MASS}$ colour efficiently
separates stars and galaxies. Based on these results, they produced a galaxy catalogue containing 2.4 million objects with  an estimated star contamination of 1.2\% and a galaxy completeness of 70\%. The separation was only made for stars and galaxies, with no information regarding quasars. A limitation of a WISE -- 2MASS cross-match is the much smaller depth of the latter with respect to the former. Most of the 2MASS galaxies are located within $z<0.2$ \citep{2MPZ,Rahman2MASS}, while WISE extends well beyond this, detecting $L_*$ galaxies at $z\sim0.5$ \citep{Jarrett16}.

Using only photometric information from WISE,  \cite{Kurcz16} employed SVM and attempted classifying all unconfused WISE sources brighter than $W1<16$ into three classes: stars, galaxies, and quasars. This led to identification of 220 million candidate stars, 45 million candidate galaxies, and 6 million candidate QSOs. The latter sample is however, significantly contaminated with what was interpreted  as a possibly very local foreground,  such as asteroids or zodiacal light.

The present paper is laid out as follows: the data are described in Sect.~\ref{Sec: Data},
Sect.~\ref{Sec:classifiaction_method} explains the principles of the support vector machine-learning algorithm and introduces the training sample used here,  and
in Sect.~\ref{Sec:classification_performance} we present various tests that allowed us to quantify the performance of the SVM algorithm. 
Section~\ref{Sec: Final classification} contains the description and properties of the final galaxy catalogue, as well as a comparison with the results of \tcb{B16} . 
In Sect.~\ref{Sec:summary} we summarise our analysis.

\section{Data: WISE and SuperCOSMOS}
\label{Sec: Data}

The catalogues used to construct the main photometric dataset, WISE and SuperCOSMOS, are comprehensively described in \cite{2MPZ} and \tcb{B16}.  Here we briefly summarize them and the preselections applied for the purpose of this project. 
They are practically equivalent to those from the latter paper; for more details see Sect. 2 of \tcb{B16}, and in particular Table 1 and Figs.\ 1--3 therein.

\subsection{WISE}
\label{Sec: WISE}

The Wide-field Infrared Survey Explorer (WISE; \citealt{WISE}), a NASA space-based mission, surveyed the entire sky in four mid-infrared (IR) bands: 3.4, 4.6, 12, and 23 $\mu$m ($W1$ -- $W4,$ respectively). 
Here we used its second full-sky release, the AllWISE dataset\footnote{Available for download from IRSA at \url{http://irsa.ipac.caltech.edu}.} \citep{AllWISE}, combining data from the cryogenic and post-cryogenic survey phases. 
It includes almost 750 million sources with signal-to-noise (S/N) ratio $\geq5$ in at least one of the bands, and its averaged 95\% completeness in unconfused areas is $W1\lesssim 17.1$, $W2\lesssim 15.7$, $W3\lesssim 11.5,$ and $W4\lesssim 7.7$ in Vega magnitudes, with variable coverage, however, that is highest at ecliptic poles and lowest near the ecliptic, especially in stripes resulting from Moon avoidance manoeuvres. 

Our WISE preselection required sources with S/N ratios higher than 2 in the $W1$ and $W2$ bands, and that they were not obvious artefacts (\ttt{cc\_flags[1,2]$\neq$`DPHO'}); this yielded 603 million detections over the whole sky. Owing to the low survey resolution ($\sim6''$), the immensely crowded Galactic Plane and Bulge are entirely dominated by stellar blends, and 
extracting particularly extragalactic information is practically impossible (extinction is not such a problem in the WISE passbands unless in very high-extinction regions, however). 
We therefore focused on the 83\% of the sky available at $|b|>10\degree$, which reduced the sample to about 460 million objects. 

The depth of WISE observations is position dependent because
of the scanning strategy, and it is highest at the ecliptic poles \citep{Jarrett11} and lowest at the ecliptic\footnote{\url{http://wise2.ipac.caltech.edu/docs/release/allwise/expsup/sec2\_2.html}}. The 95\% completeness limit of AllWISE over large swaths of unconfused sky is $W1=17.1$ mag (Vega)\footnote{\url{http://wise2.ipac.caltech.edu/docs/release/allwise/expsup/sec2_4a.html}} , and we independently verified by analysing the WISE source distribution in $W1$ magnitude bins that adopting a flux limit of $W1<17$ leads to a relatively uniform selection as far as instrumentally driven artefacts are concerned.  This cut  gave a final input WISE dataset of about 343 million sources at $|b|>10\degree$. 
At the bright end, this sample is dominated by stars even at high Galactic latitudes \citep{Jarrett11,Jarrett16}, and we estimate about 100 million of the WISE  sources, mostly faint, to be galaxies and quasars \citep[\tcb{B16};][]{Kurcz16}, the remainder are of stellar nature.

The WISE database currently does not offer reliable (e.g.\ isophotal) aperture photometry for the resolved sources and they are not even identified therein (except for fewer than 500,000 cross-matches with the 2MASS Extended Source Catalogue). 
We therefore used the \ttt{w?mpro} magnitudes (where the question
mark
stands for the channel number), which are based on point spread
function profile-fit measurements. The only proxy for morphological properties that we adopted here from the database is given by circular aperture measurements performed on the sources within a series of fixed radii. These were obtained without any contamination removal or compensation for missing pixels, however. 
In particular, as in \cite{2MPZ} and \cite{Kurcz16}, in the classification procedure we used a differential measure (a concentration parameter), that is  defined as 
\begin{equation}
\mtt{w1mag13} = \mtt{w1mag\_1} - \mtt{w1mag\_3}\;,
\end{equation}
where $\mtt{w1mag\_1}$ and $\mtt{w1mag\_3}$ were measured in fixed circular apertures of radii of $5.5''$ and $11''$, respectively. The \ttt{w1mag13} parameter is expected to have different distributions for point and resolved sources, which indeed is the case, as we verified against SDSS spectroscopic data described in Sect.\ \ref{Sec:Training sample}.

We note that of the four available WISE bands we did not employ the longest wavelength $W4$ because of its very low sensitivity, which leads to an overwhelming number of non-detections. In addition, whenever $W3$ was used, all the sources with $\mtt{w3snr}<2$ (upper limits and non-detections, which together dominate the $W3$ channel in our sample), were artificially dimmed by $+0.75$ mag to statistically compensate for their overestimated fluxes, which we determined to be an appropriate average correction (see the Appendix for details). Possible errors associated with this procedure are not important for our final catalogue, however,
as we finally did not employ the 12 $\mu$m passband for the overall classification because of the photometry issues, although it does bring some improvement (cf.\ Sect.\ \ref{Sec: is W3 needed}). The $W3$ information was used only in the test phase.

\vspace{1mm}

\subsection{SuperCOSMOS}
\label{Sec: SuperCOSMOS}
The SuperCOSMOS Sky Survey \citep[SCOS,][]{SCOS3,SCOS2,SCOS1} consists of digitized photographs in three bands, $B,R,I$, obtained through automated scanning of source plates from the United Kingdom Schmidt Telescope (UKST) in the south and the Palomar Observatory Sky Survey-II (POSS-II) in the north. The observations were conducted in the last decades of the twentieth century. 
The data are publicly available from the SuperCOSMOS Science Archive\footnote{\url{http://surveys.roe.ac.uk/ssa/}}, with photometric, morphological, and quality information for 1.9 billion sources.

SCOS provides source classification flags in each of the three bands, as well as a combined one, \ttt{meanClass}, which is equal to 1 if the source is resolved, 2 if it is unresolved, 3 if it
is unclassifiable, and 4 if it is likely noise \citep{SCOS2}, the two latter cases comprising a negligible fraction ($\ll 1\%$) of all the sources. 
The derived catalogue of extended sources was accurately calibrated all-sky using SDSS photometry in the relevant areas, and the calibration was extended over the remaining sky by matching plate overlaps and by using the average colour between the optical and 2MASS~$J$ bands \citep{Peacock16}.
This was not the case for the point sources, however, which very much limits their applicability for uniform source selection. 
In \tcb{B16} only the SCOS sources with $\mtt{meanClass}=1$ were used to obtain a \WISC\ galaxy sample that was further scourged
of residual quasars and stars. 
In the present paper we followed this preselection, but we recall that only a part of these SCOS sources are in fact extragalactic. 
Especially at low Galactic latitudes, this extended source catalogue is dominated by blends of stars with other stars and with extragalactic objects. 
The remaining SCOS preselections are also the same as in \tcb{B16} and earlier in \cite{2MPZ}: objects need to be properly detected with aperture photometry in the $B$ and $R$ bands\footnote{We did not use the third SCOS band, $I$, as it is too shallow.} (\ttt{gCorMagB} and \ttt{gCorMagR2} not null; quality flags \ttt{qualB} and $\mathtt{qualR2}<2048$, meaning no strong warnings or severe defects, \citealt{SCOS2}). 

The publicly available catalogue was supplemented with additional data in corners of the photographic plates that were missing from the original dataset because of so-called step-wedges \citep{SCOS2}. This mostly affects low declinations. 
The $B$ and $R$ magnitudes were additionally calibrated between the north and the south (the split being at $\delta_{1950}=2.5\degree$) to compensate for differences between effective passbands of UKST and POSS-II; see \cite{Peacock16} for details.

To preserve the all-sky photometric reliability and to mitigate problems with catalogue depth that varies from plate to plate, two flux limits were applied to the SCOS dataset, $B<21$ and $R<19.5$ (AB-like, extinction corrected). 
As already mentioned, we did not use the Galactic Plane strip of $|b|<10\degree$, where blending and high extinction make SCOS photometry unreliable. 
The resulting SCOS catalogue of extended sources outside of the Galactic Plane includes over 85 million objects.

\subsection{Cross-matched WISE $\times$ SuperCOSMOS photometric sample}

The two photometric catalogues were paired using a matching radius of $2"$. 
The resulting flux-limited cross-matched sample at $|b|>10\degree$ contains almost 48 million sources. This number includes WISE sources supplemented from an earlier cryogenic phase of observations \citep[`All-Sky',][]{WISEAllSky}, to fill in strips of missing data centred on ecliptic longitudes of $\lambda \sim 50\degree$ and $\lambda \sim 235\degree$.  

All the magnitudes were extinction corrected using the \cite{SFD} maps throughout and applying the following extinction coefficients, derived from the \cite{SF11} re-calibration (\tcb{B16}): $A_B = 3.44$, $A_R = 2.23$, $A_{W1} = 0.169$, and $A_{W2} = 0.130$. 
The usage of de-redenned magnitudes also for stars is motivated by the fact that we focus on extragalactic sources, the stellar ones being contamination for our applications. 
As these corrections are often significant in the optical, neglecting them would lead to considerable biases in the final galaxy catalogue. Still, we did not use the areas of very high extinction, $E(B-V)>0.25$, which have almost no training data for classification, and the photometry especially in the optical is problematic. This cut, removing another 7.2 million sources, is the same as applied in \tcb{B16}, where the appropriate threshold was determined through an analysis of spurious under- and overdensities in \WISC\ source distribution.

\begin{figure*}
    \includegraphics[width=.5\textwidth]{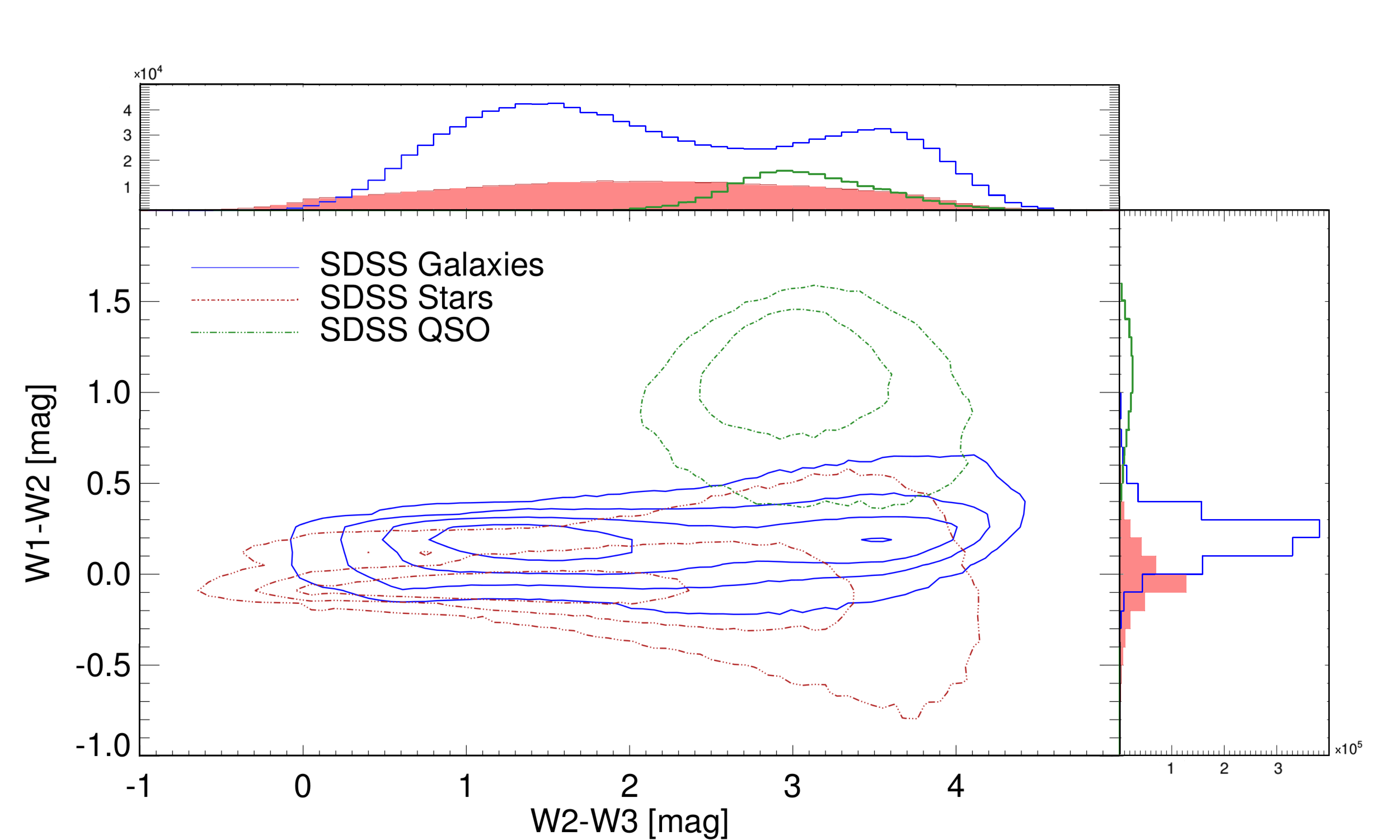}
    \includegraphics[width=.5\textwidth]{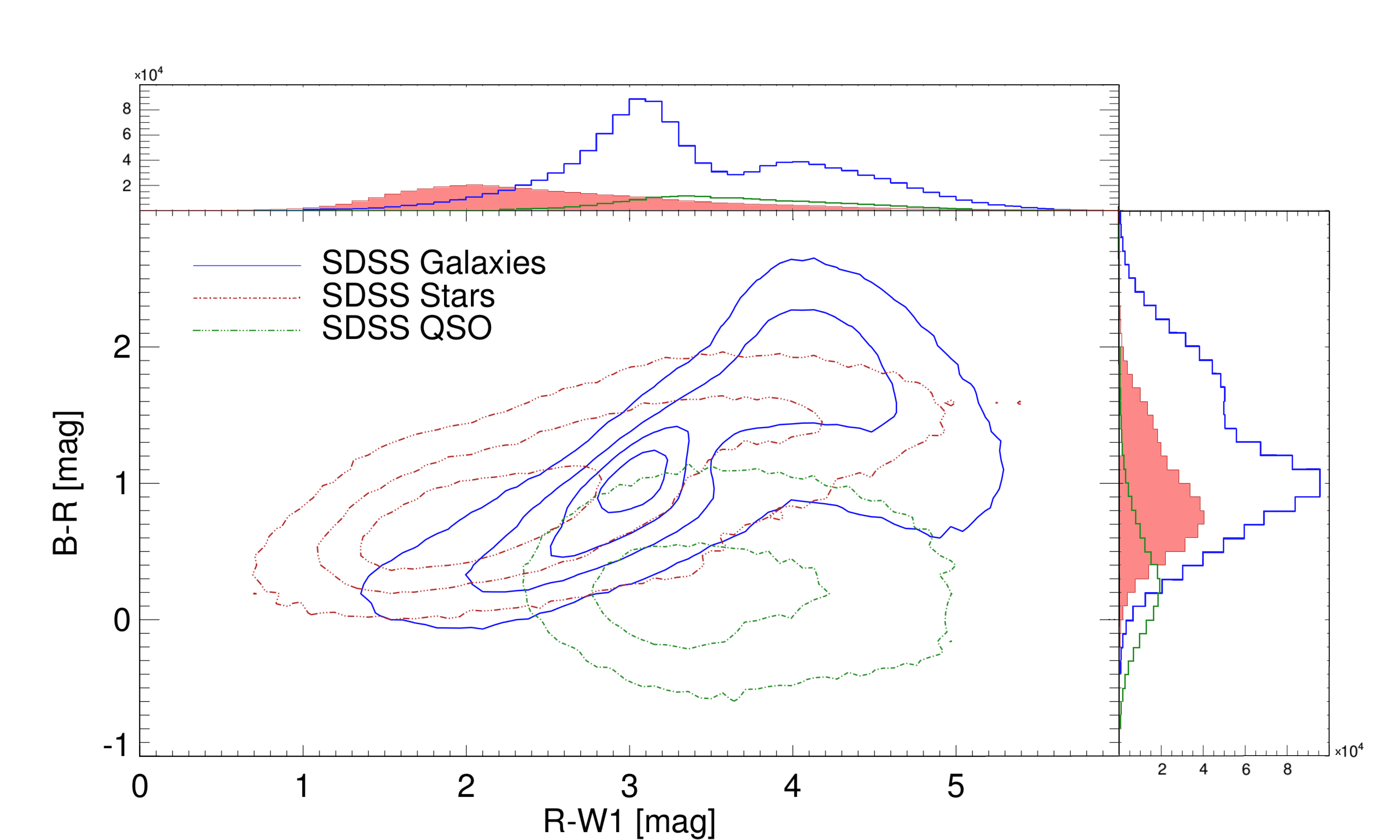}
    
        \caption {Colour-colour diagrams for galaxies, stars, and quasars from a cross-match of WISE $\times$ SuperCOSMOS with SDSS spectroscopic data.  \textit{Left panel}: WISE colours only; \textit{right panel}: WISE and SCOS colours. Blue contours correspond to galaxies, red contours represent stars, and green contours illustrate quasars.}
        \label{fig:W3CC}
\end{figure*}

Figure \ref{fig:W3CC} shows why using an automatic classifier rather than simple colour cuts is more suitable to separate galaxies from stars and quasars in the \WISC\ sample. The diagrams illustrate distributions of three source types (galaxies, quasars, and stars) on two colour-colour (c-c) planes. Source identifications come from the \WISCS cross-match described in detail in Sect.\ \ref{Sec:Training sample}. In the left panel we show the $W2-W3$ vs.\ $W1-W2$ c-c plane, which is often used for object separation in WISE \citep[e.g.][]{Jarrett11,Ferraro15}. The plot shows that it is challenging to find simple cuts in these parameters that would maximise both the completeness and the purity of the resulting samples. While we might quite well separate QSOs from other sources, for example by a $W1-W2=0.8$ cut \citep{stern12,assef13,yan13}, it is much more difficult, if possible at all, to apply a single threshold to efficiently separate galaxies from stars. The galaxy and star distributions overlap very much even if the additional colour $W2-W3$ is taken into account. The situation is similar for other colour combinations that are available from the five bands in \WISC. The right panel of Fig.\ \ref{fig:W3CC} illustrates the $R-W1$ vs.\ $B-R$ c-c plane, where the three source types also largely overlap.

\section{Classification method: support vector machines}
\label{Sec:classifiaction_method}
For the classification performed in this work we used the SVM method. SVM is a supervised learning algorithm that is a maximum-margin classifier able to determine decision planes between sets of objects with different class memberships, to establish a decision boundary by maximising the margin between the closest points of the classes (the so-called support vectors). 
Each single object is classified based on its relative position in the $n$-dimensional parameter space \citep{crisrianini00,shawe04,solarz12,malek13SVM}.

The SVM algorithm is an increasingly popular way of handling astronomical data to classify different types of objects. 
For various applications of SVM in astronomy we refer to, for instance, \cite{wozniak04,huertas08,solarz12,saglia12,malek13SVM,KoSz15,Marton2016}, and \cite{Kurcz16}.

In our case, SVM was used to build a non-linear classifier for the photometric data in the \WISC\ all-sky catalogue.
As input data we used photometric information such as magnitudes, colours, and a differential aperture magnitude (see Sect.\ \ref{Sec:classification_performance} for details). 
The input data were transformed by a kernel into a higher-dimensional feature space, where the separation between different classes is less complex than in the input parameter space. For more details see for example\ \cite{Manning:2008:IIR:1394399} or \cite{malek13SVM}; an illustrative description of how SVM classification operates is provided in \cite{Han16}.

The SVM algorithm searches for a boundary $B$ in the feature space that will separate examples from different categories by maximizing a fitness function $F$:
\begin{equation}\label{eq:fit fun}
 F=M-C\sum_i\xi_i(B,M)\;,
\end{equation}
where $M$ is the margin of the boundary, and $\xi(B,M)$ is the number of training examples violating this criterion.
The cost parameter $C$ is  a trade-off between large margins and poor classification.
Equation\ \eqref{eq:fit fun} shows that for very large $C$ each training example on the wrong side of the margin is heavily penalized.
When $C$ is small, individual $\xi_i$ penalize Eq.\ \eqref{eq:fit fun} less heavily, thus the optimal boundary may be one that misclassfies a small number of outliers. More details can be found in \cite{beaumont11}.

For our particular implementation, we used a
\texttt{C}-SVM algorithm with a Gaussian kernel
to identify three different classes of objects: galaxies, quasars, and stars, with the final aim to reliably pinpoint the galaxies. 
The Gaussian kernel function (also dubbed radial basic) is defined as
\begin{equation}
k(x_i,y_j)=exp(-\gamma||x_i-x_j||^2),
\end{equation}
where $||x_i-x_j||$ is the Euclidean distance between feature vectors in the input space.  The parameter $\gamma$ 
 is related to the breadth of the Gaussian distribution, $\sigma$, namely  $\gamma$=$1/(2 {\sigma}^2)$, and determines the topology of the decision surface. Too high a value of $\gamma$ sets a complicated decision boundary, while too low $\gamma$ can give a decision surface that is too simple, which might cause misclassifications.
The classifier is trained using a subset of input data for which class identifications are known. In our case, the training set was derived from SDSS DR12 spectroscopic data matched with \WISC\ (see Sect.\ \ref{Sec:Training sample}).

The two parameters, $C$ and $\gamma$, were optimized based on the training set through a grid search and $N$-fold cross-validation; we used $N=10$, in which case the training data
were split into ten equal sets, and the classifier was trained on nine of them. Then the classifier was tested against the remaining tenth subset (the so-called self-check or validation). This test was repeated ten times with a different subset removed for each training run. The classification accuracy was then calculated by averaging over the ten runs. The same method was used in \cite{solarz12} and \cite{malek13SVM}, where a more detailed description of this process is provided. 
Moreover, an additional test sample was used (data with known classification, but not used for training) for an independent check of the classifier performance.

In the test phase of our study, we used only the discrete classes assigned by SVM to each source. For the final classification, however, we decided to employ the full probability distributions for each class. This allowed us to examine the cases of problematic classification where the probabilities that a source belongs to two or three classes were roughly equal. This is discussed
in detail in Sect.\ \ref{Sec: Final classification}.

For our analysis we used LIBSVM6 \citep{chang11}, integrated software for support vector classification, which allows for multiclass identifications. 
We also employed \ttt{R}, a free software environment for statistical computing and graphics, with the \ttt{e1071} interface \citep{meyer01} package installed.

\vspace{1.1cm}

\subsection{Training sample: SDSS DR12 spectroscopic data}
\label{Sec:Training sample}

A well-chosen training sample is crucial for the SVM method, because the classifier is tuned based on the properties of this sample: the $C$ and $\gamma$ parameters are estimated and the hyperplane between classes  is determined. This means that a representative sample of sources, with known properties that we wish to identify, is essential. 
In our specific case of a catalogue including $z\lesssim 0.5$ galaxies \citep[][\tcb{B16}]{2MPZ} as well as stars and higher-redshift quasars, such a training set requires good-quality and high-reliability pre-classification using spectroscopic measurements. For this reason, for training and testing purposes we chose to employ the spectroscopic sample from the Sloan Digital Sky Survey Data Release 12 (SDSS DR12, \citealt{SDSS.DR12}) cross-matched with the \WISC\ dataset defined above. 

The SDSS is a multi-filter imaging and spectroscopic survey, and  its DR12, used for our analysis, includes dedicated star, galaxy, and quasar surveys. 
These samples are shallower than the imaging part of the SDSS, but they are available with high reliability only from spectra \citep{Bolton12}: the photometric classification of SDSS was only based on source morphology (resolved vs.\ point-like, \citealt{SDSSphoto}). 
SDSS DR12 contains almost 3.9 million spectroscopic sources, of which 61\% were identified as galaxies, 22\% as stars, and the remaining 16\% as quasars/AGNs (SDSS class `\ttt{QSO}'). 
To avoid unreliable spectroscopic measurements and hence problematic classification, we used additional information on the redshift from the SDSS database as a quality determinant: the \ttt{zWarning} flag and the relative error in redshift (radial velocity for stars) defined as $\Delta z = z_\mrm{err}/z$, where $z_\mrm{err}$ is the database value. 
Only the sources with $\mtt{zWarning}=0$ were used throughout, with the additional conditions of $\Delta z < 0.1$ for galaxies and quasars, and $\Delta z < 1$ for stars. 

Pairing these sources with our \WISC\ flux-limited catalogue within $1"$  matching radius resulted in over 1 million common objects, 95\% of which were galaxies, 2\% were stars, and 3\% were quasars. 
Clearly, the stars and most of the quasars are point sources and should not be resolved. 
That we identified over 50,000 of them in the cross-match of SDSS with the \WISC\ extended source catalogue reflects the susceptibility of SCOS morphological classification (the \texttt{meanClass} flag) to blending, which mimics resolved sources (\tcb{B16}). 
The main purpose of the present study is to reliably filter out such sources from the galaxy catalogue we aim to produce. 

\section{SVM classification performance}
\label{Sec:classification_performance}

This section describes various tests made using the SDSS-based training sample, which allowed us to quantify the performance of the SVM algorithm in view of the final classification of the entire catalogue. 
 
To check the classification efficiency, to calculate the dependence on different parameters, and to perform the final classification, the following procedure was used: (1) as galaxy properties change with magnitude, each sample (training and test sets, final catalogue) were divided into five $W1$ magnitude bins ($W1< 13$, $ 13 \leq W1 < 14 $, $14 \leq W1 < 15$ , $15\leq W1 < 16, $ and $16 \leq W1 < 17 $); (2) five SVM algorithms separately tuned for these bins were used to classify galaxies, stars, and quasars; and
(3) five SVM outputs were merged and treated as one final output.  
We verified that there is no evidence for an inconsistency between different $W1$ magnitude bins.

 \begin{figure}
  \includegraphics[width=0.5\textwidth]{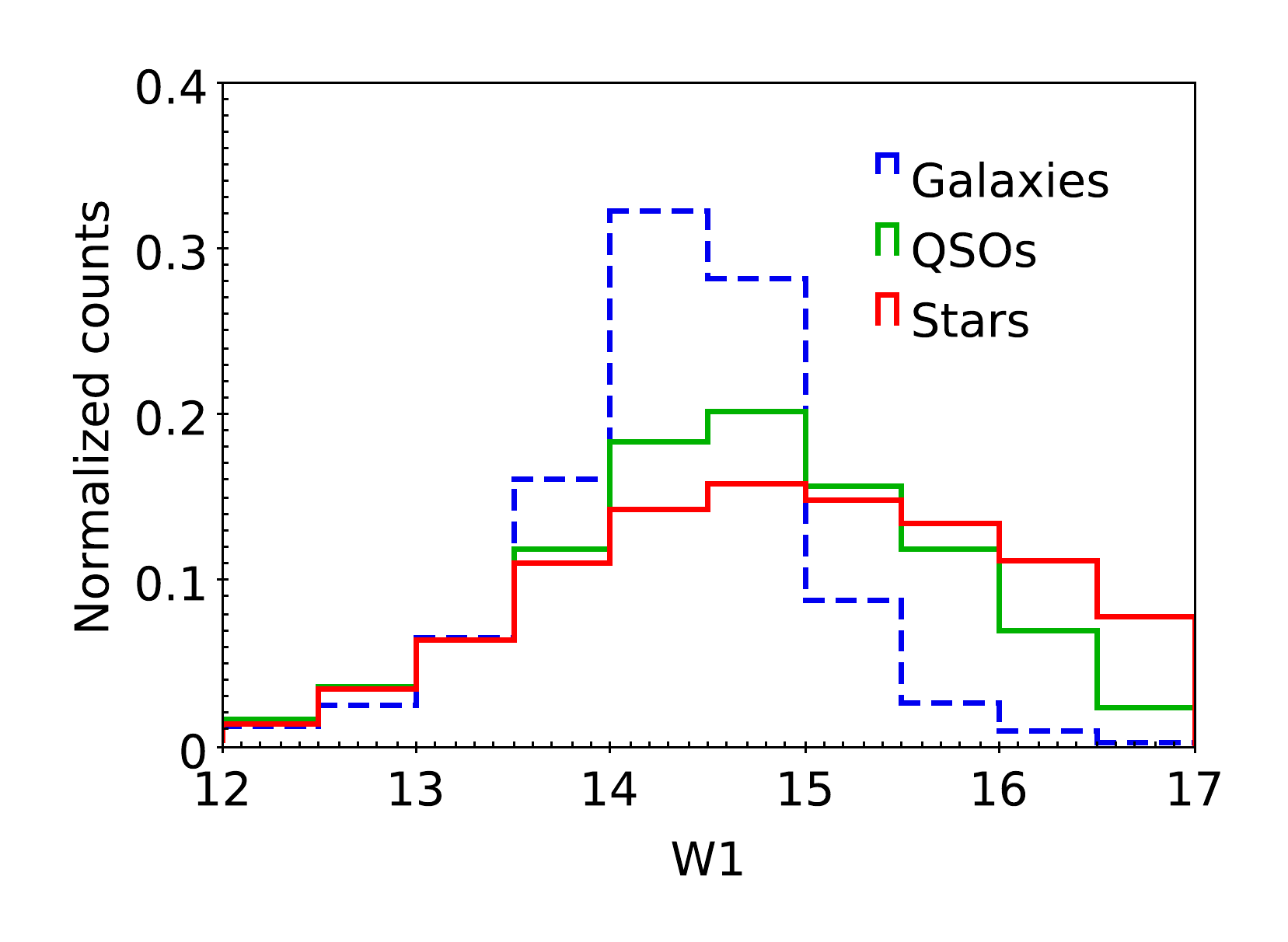}
        \caption{Normalised number counts of $W1$ magnitudes in the \WISCS cross-matched sample for stars, galaxies, and quasars.}
        \label{fig:histo_wszystkieSDSS}
\end{figure} 

Figure \ref{fig:histo_wszystkieSDSS} shows normalized $W1$ magnitude distributions for the three types of sources in the \WISCS cross-match. 
As was shown in \cite{Kurcz16}, the derived classification statistics  depends on the number of training objects, but the classifier stabilizes for subsamples of 3000 objects in each class. 
In our case, however, at both the bright ($W1<13$) and the faint end ($W1>16$), we did not have large enough numbers of sources 
to select randomly 3000 objects of each class from the input sets to build the training sample. 
In particular, we needed to use all the stars and quasars from these bins for the training and tests.  
For this reason, our training samples consist of different numbers of objects in each $W1$ bin: 1000, 4000, 4000, 5000, and 2600 of each type for the $12<W1<13$,   $13<W1<14$, $14<W1<15$, $15<W1<16,$ and $16<W1<17$ mag bins, respectively.

\begin{figure}
        \resizebox{\hsize}{!}{\includegraphics[scale=0.5,clip]{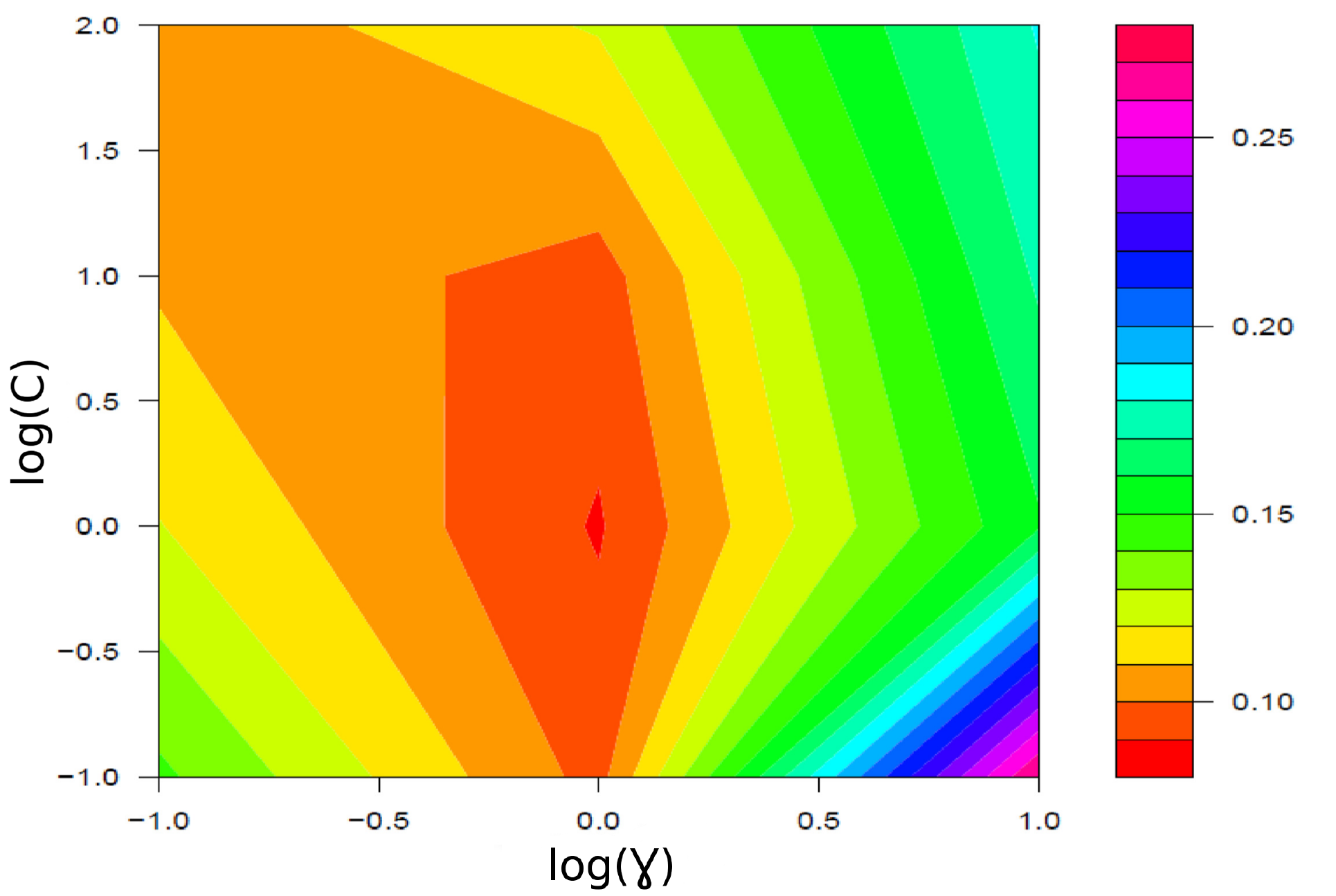}}
    \centering
        \caption{\label{fig:svmCgamma} Example of the $C-\gamma$ plane obtained from one of the five \WISC\ C-SVM classifiers. 
        The mean misclassification rate (colour bar) as a function of the $C$ and $\gamma$ parameters was estimated through ten-fold cross-validation for each pair of the two parameters. 
        The lower the misclassification rate, the better the performance of the SVM algorithm.}
\end{figure}

The first step of the tests was to determine the optimal $C$ and $\gamma$ parameters for the five $W1$
bins , therefore we tuned five different \texttt{C}-SVM classifiers for our purpose. 
Figure~\ref{fig:svmCgamma} illustrates an exemplary grid search for one of the classifiers. 
The colours code the mean misclassification rate for given combinations of  $\gamma$ and $C$; the lower the rate, the better the performance of the SVM algorithm.
Here the misclassification rate is defined for each magnitude bin as the complement to the total accuracy (TA), the latter being the mean of accuracies $A_i$ for individual validation iterations:
\begin{equation}\label{TotalAccuracy}
\mrm{TA}=\frac{1}{10}\sum_{i=1}^{10} A_{i}\;.
\end{equation}
The accuracy for a given iteration is defined as 
\begin{equation}\label{accuracy_single}
A_i=\rm\frac{TG+TQ+TS}{TG+TQ+TS+FG+FQ+FS}\;.
\end{equation}
The components of this equation are true galaxies (TG), quasars (TQ) and stars (TS) from the training sample, properly classified as galaxies, quasars, and stars, respectively; and 
false galaxies (FG), which are real quasars or stars misclassified as galaxies, with false quasars (FQ) and false stars (FS) defined in a similar manner. 

To further compare the performance of different classifiers, we calculated the following measures, as defined by \cite{soumagnac13}: 
completeness ($c$), contamination ($f$), and purity ($p$) for galaxy, star, and quasar samples. 
We used the following equations (here for galaxies):

\begin{equation}
{c_g=\rm{\frac{TG}{TG+FGS+FGQ}}},
\label{gc}
\end{equation}
\begin{equation}
{f_g=\rm{\frac{FSG+FQG}{TG+FSG+FQG}}},
\label{gf}
\end{equation}
\begin{equation}
p_g=\rm{1-}f_g=\rm{\frac{TG}{TG+FSG+FQG}},
\end{equation}
where  FGS and FGQ stand for galaxies misclassified as stars and quasars, and FSG, FQG are stars and quasars misclassified as galaxies. 
Definitions for stars and quasars follow in an analogous way. 
The accuracy for an individual class of objects is defined in the same way as the purity.

\subsection{Usefulness of the $W3$ passband for the classification}
\label{Sec: is W3 needed}

We tested two classifiers for the separation between galaxies, quasars, and stars: one with five and the other with six parameters. 
These were $W1$ magnitude, $W1-W2$ colour, $R-W1$ colour, $B-R$ colour, and the \ttt{w1mag13} differential aperture magnitude for the $W1$ channel. 
The sixth parameter in the tests was the $W3$ magnitude, which
is often used in WISE-based source classifications \citep[e.g.][]{KoSz15,Ferraro15}, following the considerations of \citet{WISE}, for example, that different types of sources occupy different regions of the $W1-W2$ vs.\ $W2-W3$ colour plane. 
However, as Fig.\ \ref{fig:W3CC} shows, this idealized picture becomes more complicated for actual observations, and we decided to test how much the $W3$ passband from WISE improves the automatic classification.
We note that to avoid biases for overestimated fluxes, a recalibration of the $W3$ upper limits was necessary, as discussed in Sect.\ \ref{Sec: WISE} and detailed in the Appendix; 
this did not prevent very low S/N $W3$ measurements (which dominate the sample) from introducing possible confusion, however.

\begin{table}[h!]
\begin{center} 
\caption{ Comparison of the performance for two classifiers: one using five parameters 
($W1$, $W1-W2$, $R-W1$, $B-R$, \ttt{w1mag13}) and the other adding $W3$ as the sixth parameter. 
TA = total accuracy; $c$ = completeness,  and $p$ = purity, all calculated as a weighted arithmetic mean for all five $W1$~bins.
}
\begin{tabular}{c|c|c|c|c|}
\cline{2-5}
    &\multicolumn{2}{c|}{5D classifier} &\multicolumn{2}{c|}{6D classifier}\\ \cline{2-5}
    &$c$ [\%] & $p$ [\%]& $c$ [\%] & $p$ [\%] \\  \hline \hline
galaxies &  90.3  & 89.9& 96.8 & 96.7  \\ 
quasars &  95.1  & 92.2 & 98.1 & 98.6  \\
stars   &  90.0  & 88.5 & 96.9 & 98.1  \\ 
 \hline \hline
TA  & \multicolumn{2}{c|}{91.8 \%} & \multicolumn{2}{c}{97.3\%} 
\label{tab:2class}
\end{tabular} 
\end{center}
\end{table}

The results for the two classifiers are  summarized in Table~\ref{tab:2class}. 
The accuracy, completeness, and purity for both cases are very high.  
The contamination levels rarely exceed $10\%$ and  $5\%$ for
the 5D and 6D classifier, respectively. 
The 6D classifier clearly provides better results in all the calculated metrics, which shows that the availability of the $W3$ band allows for (possibly considerable) improvement in the classification. However, the low detection rate in this band (only 30\% of our sources have $\mtt{w3snr}>2$) and the large variations in sensitivity on the sky\footnote{See e.g.\ \url{http://wise2.ipac.caltech.edu/docs/release/allsky/expsup/sec6\_2.html}.} mean that using this band might introduce biases into the final catalogue.
Based on these
 considerations,
together with the fact that each new classification parameter extends computation time, we decided not to use the $W3$ passband for the final classification. 

\begin{figure*}
        \centering
        \begin{subfigure}[b]{0.5\textwidth}
                \includegraphics[width=\textwidth, height=0.3\textheight]{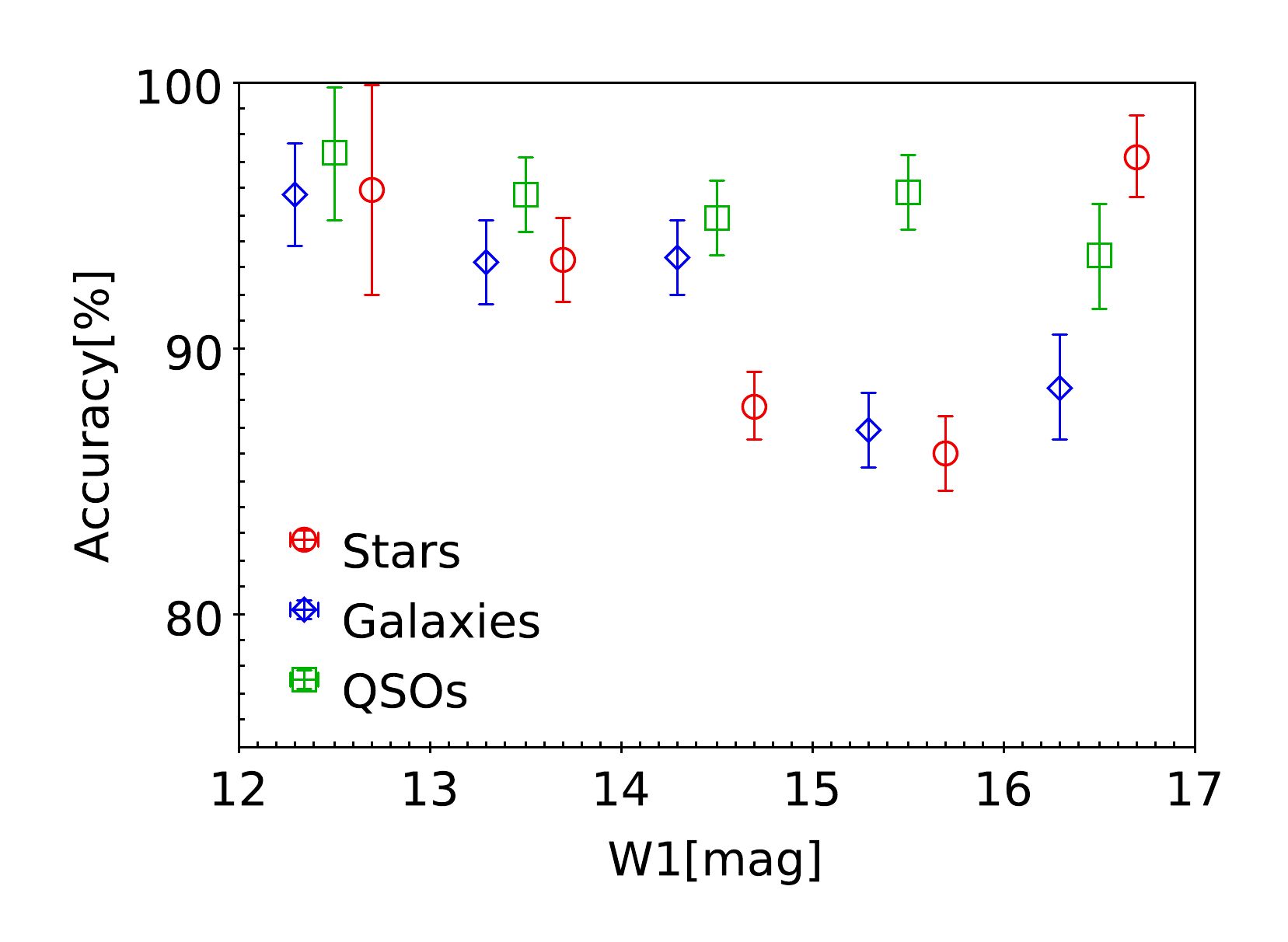}
                \caption{Self-check}
                \label{fig:gal_qso_star_acc_TS}
        \end{subfigure}%
      \begin{subfigure}[b]{0.5\textwidth}
                \includegraphics[width=\textwidth, height=0.3\textheight]{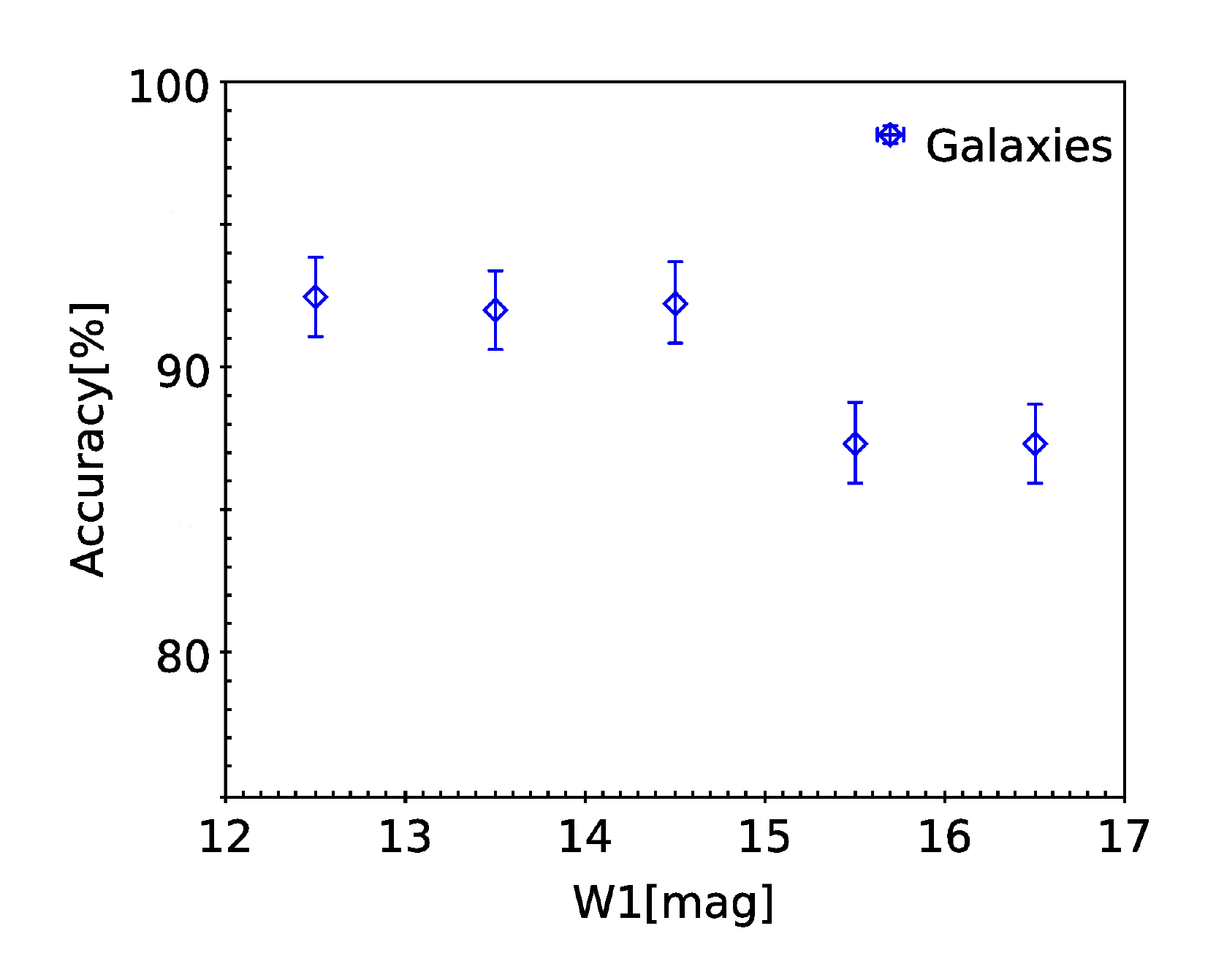}
                \caption{Test sample}
                \label{fig:gal_qso_star_acc_test}
        \end{subfigure}%
        \caption{Accuracy of the 5D classifier as a function of the limiting $W1$ magnitude for the three types of classified sources, for the self-check (\textit{left panel}) and the test sample (\textit{right panel}).
        Blue diamonds correspond to galaxies, green squares to quasars, and red circles to stars. The points were shifted horizontally for clarity. }
        \label{fig:W1dependence}
\end{figure*}

\subsection{General performance of the classifier}
\label{Sec: general SVM performance}
To quantify the general performance of the set of five classifiers tuned for different $W1$ magnitude bins, we analysed the final results (merged output catalogues from different $W1$ bins) of the self-check and the test sample. 
As the test sample we randomly chose 5,000 galaxies from the same \WISCS catalogue, independent of the training sample. The total accuracy, calculated over the galaxy test sample, was equal to 92.5\%. It was not possible to perform the same analysis for stars and quasars because for the brightest and faintest bins, all of them were used to build the training sample.
\begin{figure*}
        \centering
        \begin{subfigure}[b]{0.5\textwidth}
                \includegraphics[width=\textwidth, height=0.3\textheight]{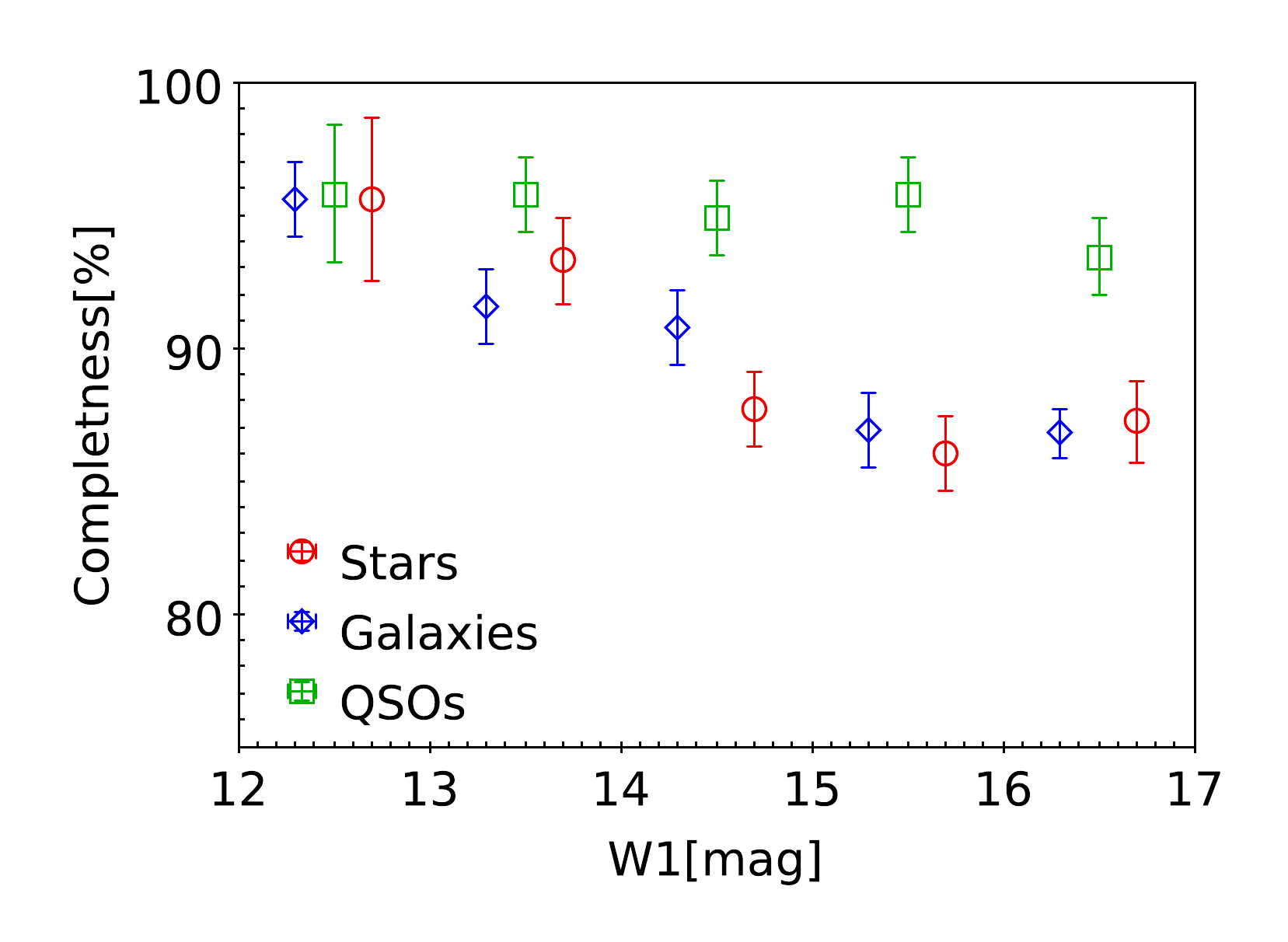}
                \caption{completeness}
                \label{fig:completness_W1}
        \end{subfigure}%
      \begin{subfigure}[b]{0.5\textwidth}
                \includegraphics[width=\textwidth, height=0.3\textheight]{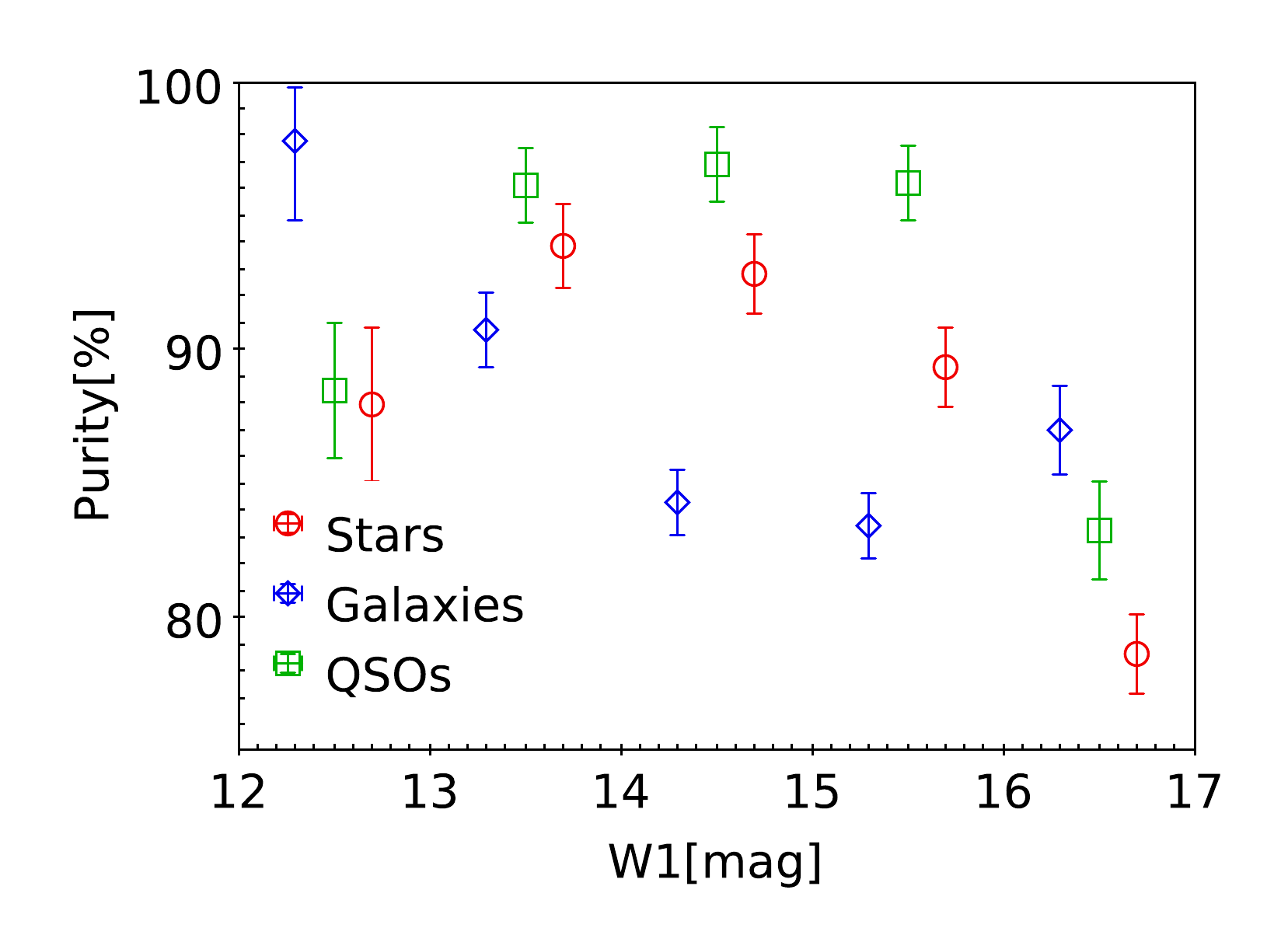}
                \caption{purity}
                \label{fig:purity_W1}
        \end{subfigure}%
        \caption{
        Completeness and  purity for galaxies, stars, and quasars as a function of the $W1$ magnitude for the 5D classifier. 
        These results refer to the self-check.  }
        \label{fig:comp_purity_W1}
\end{figure*} 

\subsubsection{Dependence on the $W1$ magnitude}

After determining the total accuracy of the 5D classifier, we investigated its performance in more detail, starting with the dependence on the $W1$ magnitude for the three classes. 
The results are illustrated in Fig.~\ref{fig:W1dependence} for the self-check (left panel) and test samples (right panel, only for galaxies). 
In general the accuracies retain very high levels of about 90\%, but there is significant deterioration in classification quality  for faint galaxies and stars. 
This is related to the fact that beyond $W1\gtrsim 15.5$ the training set contains very few galaxies and stars. 
The misclassification of galaxies occurs mainly for objects with $W1>15$ mag, and in most cases, true galaxies are misclassified as stars. 
The accuracy for galaxies calculated for the test sample has the same dependence on $W1$ magnitude as the one derived from the self-check. 

As we show in Fig.~\ref{fig:completness_W1}, the completeness of the galaxy sample also decreases with increased $W1$. 
For galaxies in the $15 \leq W1 < 17$ mag bin, the  completeness equals $\sim 87\%$. 
This deterioration  was expected, as there are far fewer training objects in the galaxy and star samples for the faintest $W1$ bin than in the quasar sample.

We also checked the purity as a function of $W1$ for the sources classified with the 5D classifier. 
As Fig.~\ref{fig:purity_W1} shows, it is at similar levels as the completeness, although its dependence on $W1$ is somewhat different. 
In particular, for all the three classes, there is a significant decrease in purity at the faint end. Still, as far as galaxies are concerned (of main interest for the present analysis),  it stays at a reasonable level of $p>80\%$ in all the bins.

Based on the findings of this section, we conclude that the 5D classifier is stable and can be safely used for the final classification.  
In principle, using the self-check and test results, we could estimate the main statistics of the final \WISC\ galaxy catalogue. 
The caveat is, however, that the SDSS training sample may not be representative enough for the \WISC\ dataset, which can lead to biases in such assessments of the final sample quality.

In the final catalogue we will keep all the sources preselected as in Sect.\ \ref{Sec: Data}, but owing to the above considerations,  it might be preferable for more sophisticated analyses to remove the faintest sources to avoid possible misclassification.   Nevertheless, we stress that the classification accuracy for the fainter part of our galaxy sample is still satisfying as it reaches very high levels even for the faintest sources (87\% for $15<W1<16$ mag bin, and $\sim$90\% for galaxies with $W1>16$ mag for the self-check of the 5D classifier, and more than 85\% for the galaxy test sample with $W1>15$ mag).

\subsubsection{Dependence on Galactic latitude}

\begin{figure}[!t]
 \centering
 \includegraphics[width=.5\textwidth]{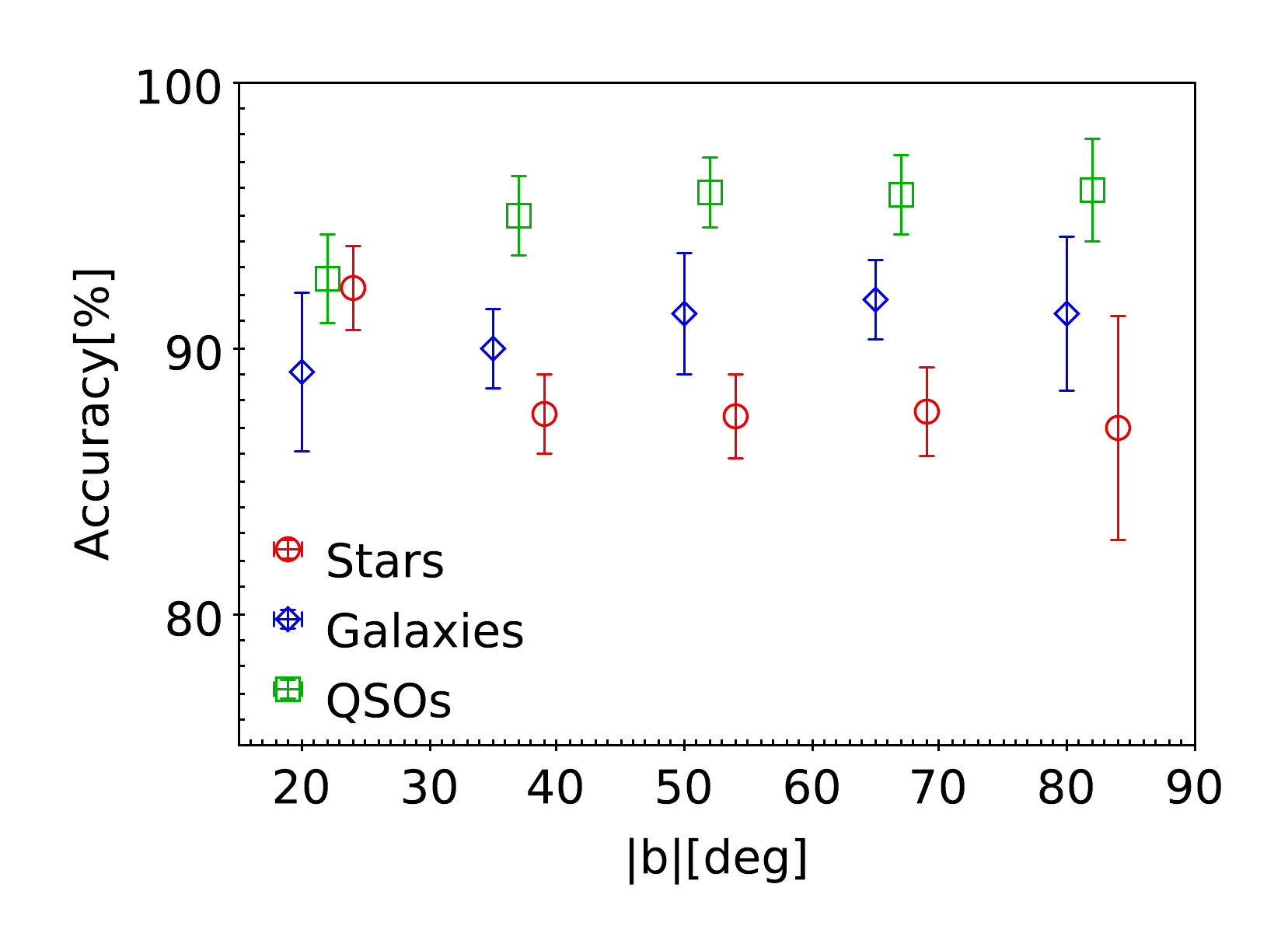}
 \caption{ Accuracy of the 5D classifier as a function of Galactic latitude for the three source classes.}
 \label{fig:accuracy_vs_b}
\end{figure}

We also checked how the accuracy of the five classifiers depends on Galactic latitude, $b$. 
We divided the training sample into six $15\degree$-wide bins in $|b|$, and calculated the accuracy for each of the bins. 
The results are shown in Fig.~\ref{fig:accuracy_vs_b}.   
For the lowest latitude bin of $|b|<15\degree$, the training sample contains practically no galaxies nor quasars, 
it was therefore not used in this test. This also means that to avoid extrapolation, this area may need to be discarded from the eventual galaxy catalogue.

\section{Results: final galaxy catalogue}
\label{Sec: Final classification}
After thoroughly verifying the performance of the SVM algorithm on the test data, we applied it to the full \WISC\ sample described in Sect. \ref{Sec: Data}. To prepare the final galaxy catalogue based on our automatic classification, we used additional information provided by SVM, namely the probabilities that the sources belong to particular classes. We also checked the catalogue for outliers in magnitude and colour space, and finally we compared it with the catalogue presented in \tcb{B16,} where simple colour cuts were employed to remove stars and quasars.

Although the SVM classifier assigns the final distinction based on discrete classes, it also provides additional information on object distance from different boundaries, which can be used as a probability for a given source to belong to a particular class. 
The probability calculated in SVM is given by the formula from \cite{platt99}, and this a posteriori probability function was implemented in the SVM kernel by \cite{lin07}. 
For classification into more than two source types, the single class probabilities are combined together to estimate final probabilities  by the pairwise coupling method (for detailed information see \citealp{wu03}).  
As our aim is to obtain a pure galaxy sample (with a strong decision value), we decided to take advantage of the full probability distributions to eliminate sources of unclear classification (located between different  classes), instead of using discrete classes alone. Initially, the galaxy candidate catalogue output by SVM (i.e.\ such that $p_\mathrm{gal}>p_\mathrm{star}$ and $p_\mathrm{gal}>p_\mathrm{QSO}$) included 
over 16.8~million sources.  
As in \cite{Kurcz16}, here we also checked whether cuts on source type probabilities might lead to an improvement in quality of the catalogue. 
Unlike in that analysis, however, in the case of \WISC\ galaxies even a cut of $p_\mathrm{gal}>0.5$ (as well as more aggressive cuts) did not lead to an increase in the purity of the sample, while it lowered its completeness. For the subsequent analysis we therefore kept all the sources flagged as galaxies by SVM. 
We note that the derived SVM probability values are made available in the \WISC\ database. This will allow users to apply their own cuts to  purify the sample (at the expense of completeness), for instance by setting maximum thresholds on $p_\mathrm{star}$ and $p_\mathrm{QSO}$, or cutting more aggressively on $p_\mathrm{gal}$.

\begin{figure*}
        \centering
        \includegraphics[width=0.99\textwidth]{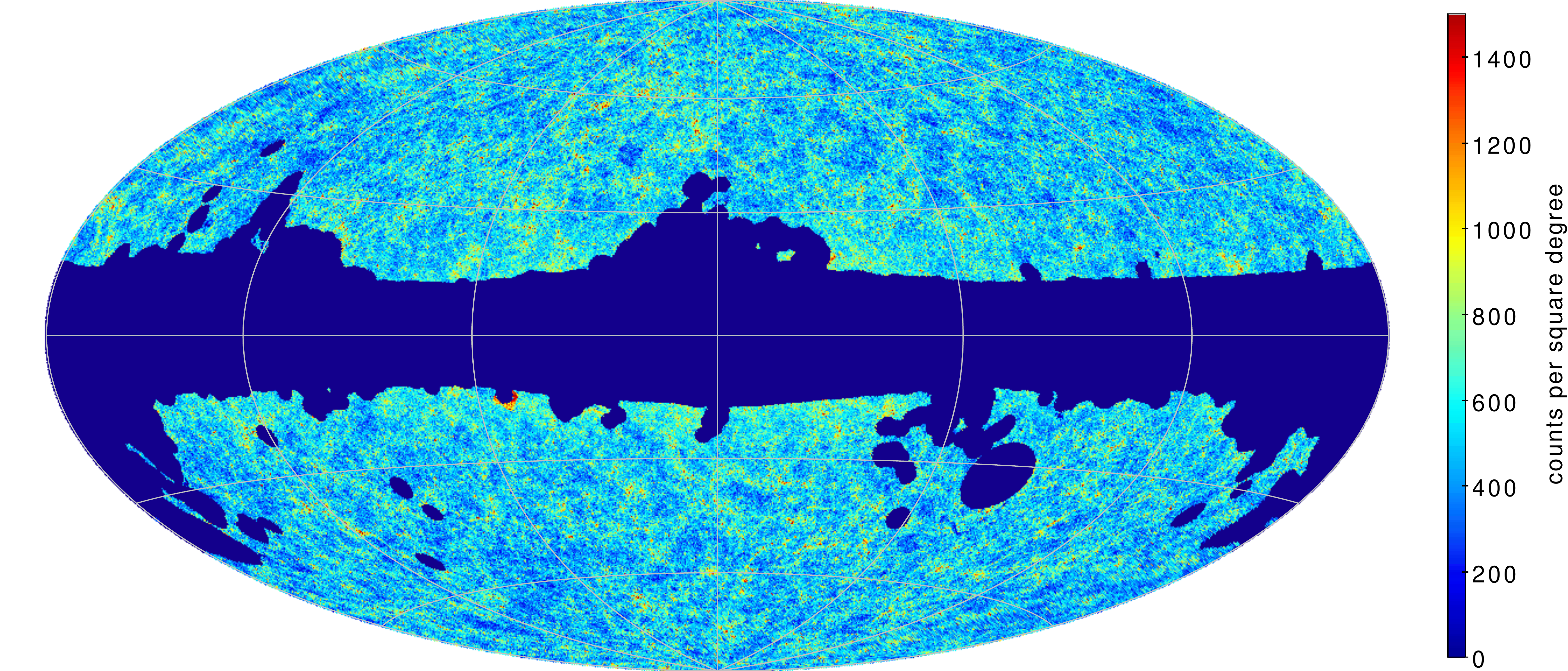}
\caption{Aitoff projection of sources identified by SVM as galaxy candidates in the \WISC\  catalogue after masking (see text for details). This plot shows 15 million objects in Galactic coordinates (with $\ell=0\degree, b=0\degree$ at the centre).}
        \label{Fig: SVM galaxies map}
\end{figure*}

The resulting catalogue was examined further for possible outliers in magnitude and colour space. Here we used the \WISCS data as the calibration to determine cases of extreme extrapolation from the training data. We found a very small number (only $\sim$1500) of sources that had colours very different from those in the calibration sample. Most of them are located near the Galactic Plane or by the Magellanic Clouds, where \WISC\ photometry is problematic because of blending. These areas need  to be masked out with a mask such as the one derived in \tcb{B16}. Applying that mask to the current catalogue, we were left with 15~million sources that are shown in Fig.~\ref{Fig: SVM galaxies map}.

\subsection{Comparison with \cite{WISC16}}

It is interesting to compare the \WISC\ galaxy catalogue derived in this paper with the dataset presented in \tcb{B16}. The parent sample used in that work was the same as ours, but galaxies were separated from stars and quasars through colour cuts. In particular, the star-galaxy separation was made through a position-dependent cut in the $W1-W2$ colour to accommodate variations in the stellar locus with the position in the Galaxy. At high Galactic latitudes the cut was $W1-W2>0$, while it was gradually increased at lower latitudes to reach $W1-W2>0.12$ by the Galactic Plane and Bulge; see Sect. 4.2 and the Appendix of \tcb{B16} for details. To this, three cuts were added to mitigate stellar contamination and blending: (i) removal of the bright
end of the sample ($W1<13.8$), which is dominated by stars on the one hand and is already sampled extragalactically by the 2MASS Photometric Redshift catalogue \citep[2MPZ,][]{2MPZ} on the other; (ii) a cutout of the Galactic Bulge reaching up to $|b|=17\degree$ at $\ell=0\degree$; and (iii) manual cutouts of the Magellanic Clouds and M31. Finally, quasars and blends thereof were removed by \tcb{B16} with colour cuts in the $(W1-W2)$ - $(R-W2)$ plane: anything with $R-W2>7.6-4(W1-W2)$ or $W1-W2>0.9$ was discarded. These cuts resulted in a dataset of 21.5 million sources; however, the sample still presented some spurious over- and underdensities in some areas, and an iterative procedure was performed to design the final mask. After the masking, the eventual \WISC\ galaxy catalogue of \tcb{B16} included 18.7 million sources  over 68\% of the sky. 

A cross-match of the SVM galaxy dataset with the dataset presented  by \tcb{B16} gives over 14.8~million common sources, which means that there are almost 2 million objects identified by SVM as galaxies that had been removed from the \tcb{B16} sample. However, for this comparison to be meaningful, we should also remove the $W1<13.8$ sources from the SVM catalogue, as well as those in the Bulge area, in the same way as in \tcb{B16}. 
These two cuts reduce the  sample generated by SVM but not by the colour cuts
to 1.3 million, which is
roughly 8\% of the original SVM galaxy dataset. 
These objects are mostly concentrated at low Galactic latitudes ($|b|<30\degree$) and around the Magellanic Clouds, that is to
say,\ in areas where the stellar blending that affects both parent catalogues has a negative impact on the photometry of extracted sources. Practically all of these objects have $W1-W2<0.12$, as expected (the upper limit of the \tcb{B16} adaptive cut), and 1~million of them are outside the \tcb{B16} mask. In general, the colours of the sources that are identified by SVM as galaxies but are absent from the \tcb{B16} catalogue are consistent with those of SDSS stars or quasars. 

Interestingly, practically no sources identified by \tcb{B16} as quasars are present in the SVM galaxy catalogue: 1300~objects that meet the QSO colour criteria mentioned above are found in the SVM dataset. 
As those colour cuts were calibrated on a comparison of SDSS QSOs and GAMA galaxies \citep{GAMA}, we conclude that our present catalogue is practically free of quasar contamination. 
This is consistent with the results from the tests presented in Sect.\ \ref{Sec: general SVM performance}.

There are significantly more sources (5.6~million after masking) in the \tcb{B16} galaxy catalogue that are absent from the SVM catalogue than the other way round. 
They are generally distributed over the entire sky, although their surface density increases towards the Galactic Plane. 
Their $W1-W2$ colour distribution is bimodal, with one peak at $W1-W2\sim0.1$ and the other at $W1-W2\sim0.5$. 
The former might indeed be stars that survived the position-dependent cut of \tcb{B16}, but were correctly classified by SVM. 
The latter are probably starburst or dusty galaxies, which our SDSS-based training sample is less sensitive to, hence they were partly misidentified by the classifier and removed from the SVM dataset; they were (correctly) kept in the \tcb{B16} sample, however.

\begin{figure}
        \centering
        \includegraphics[width=.45\textwidth]{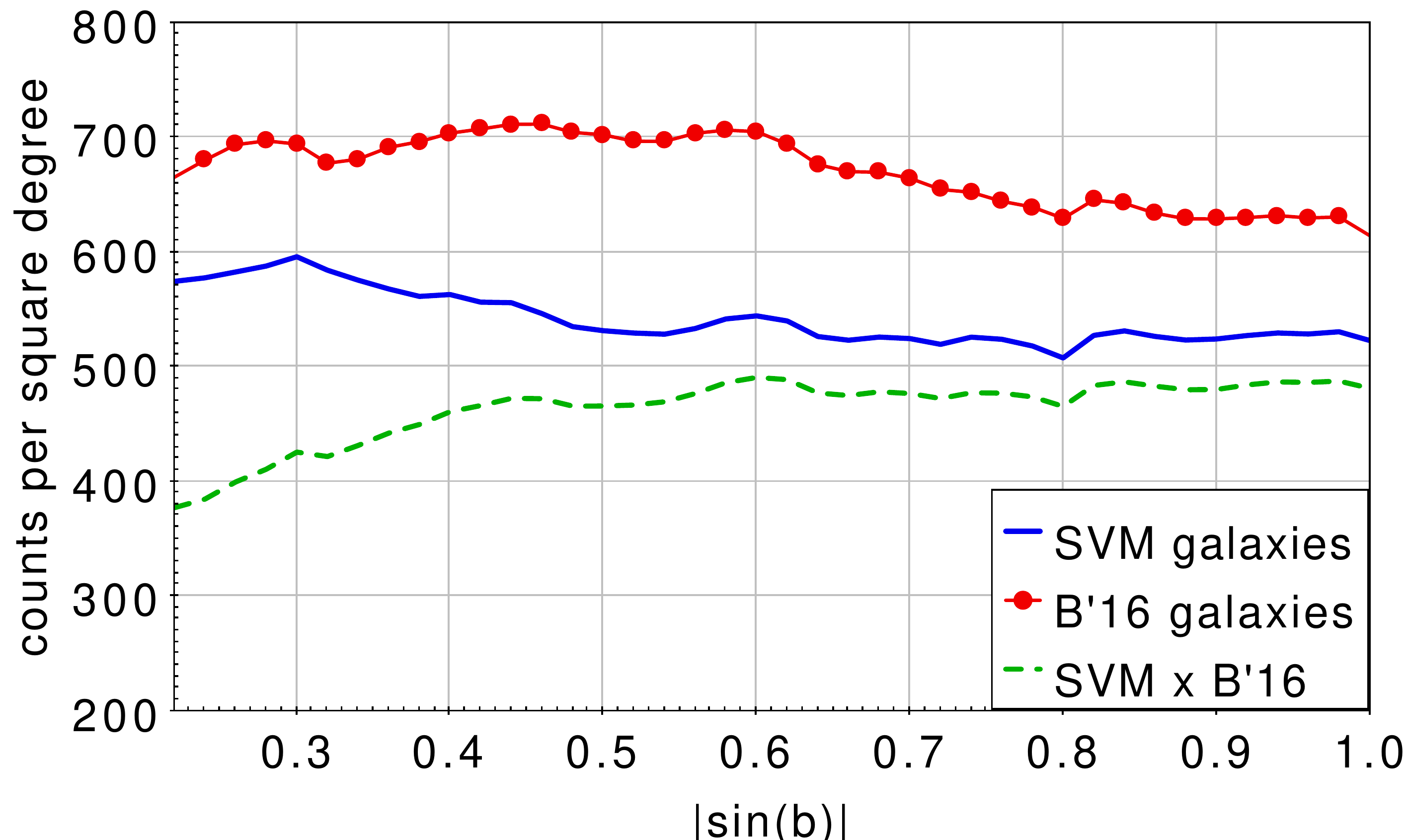}
\caption{Number counts as a function of the sine of Galactic latitude for three samples of WISE $\times$ SuperCOSMOS galaxies: selected with colour cuts by \cite{WISC16} (red dotted), identified by SVM (present work; blue solid), and common to both datasets (green dashed).}
        \label{Fig: counts abs bGal}
\end{figure}

Finally, a comparison of source counts as a function of Galactic latitude (Fig.\ \ref{Fig: counts abs bGal}) suggests that the SVM catalogue is purer than the catalogue assembled by \tcb{B16}, as the rise in the number counts with decreasing absolute latitude occurs at lower $|b|$ in the former than in the latter.  However, as \tcb{B16} estimated that their catalogue was less than $90\%$ complete at $|b|>30\degree$ ($|\sin b|>0.5$), and the absolute counts of the present dataset are lower than those of the \tcb{B16} sample even in the Galactic caps,  we conclude that the higher purity of the SVM catalogue comes at the price of lower completeness. 

\section{Summary}
\label{Sec:summary}

The \WISC\ galaxy sample is currently the largest in terms of its size and sky coverage at $z\sim0.2$, giving access to angular scales not accessible with samples such as SDSS. At the same time, it is much deeper than other all-sky datasets that are available from IRAS or 2MASS. 
Here we presented an approach to identify galaxies in the \WISC\ photometric data that is an alternative to the colour cuts applied in \cite{WISC16}. 
By using the support vector machines algorithm, trained and tested on a cross-match of spectroscopic SDSS data with \WISC, we identified about 15~million galaxy candidates over 70\% of sky. 
This number is smaller than 18.5~million obtained by \tcb{B16}, mostly because our sample is of higher purity but lower completeness than the colour-selected sample. 
The resulting source probabilities assigned by SVM are provided in the photometric redshift \WISC\ dataset released together with the publication of \tcb{B16} , available from the Wide Field Astronomy Unit, Institute for Astronomy, Edinburgh at \url{http://ssa.roe.ac.uk/WISExSCOS.html}.

We focused on galaxies because we used only extended (resolved) sources from SuperCOSMOS. 
Still, this work might be continued to obtain a more general identification of stars, galaxies, and quasars in the full \WISC\ sample. 
This would require SCOS point-source photometry to be calibrated all-sky in a similar way as the aperture-based measurements \citep{Peacock16}, however, which currently is not the case. 

Successful machine-learning galaxy identification in \WISC\ shows that a similar approach will be worthwhile for other samples based on WISE, cross-matched with forthcoming wide-angle datasets such as Pan-STARRS, SkyMapper, or VHS. 
For WISE itself, first efforts of all-sky star, galaxy, and QSO separation in that catalogue have been reported in \cite{Kurcz16}.

\begin{acknowledgements}

We are grateful to John Peacock for his useful comments. Special thanks to Mark Taylor for the TOPCAT\footnote{\url{http://www.star.bristol.ac.uk/\~mbt/topcat/}} \citep{TOPCAT} and STILTS\footnote{\url{http://www.star.bristol.ac.uk/\~mbt/stilts/}} \citep{STILTS}  software. Some of the results in this paper have been derived using the HEALPix package\footnote{\url{http://healpix.sourceforge.net/}} \citep{HEALPIX}.\\

This publication makes use of data products from the Wide-field Infrared Survey Explorer, which is a joint project of the University of California, Los Angeles, and the Jet Propulsion Laboratory/California Institute of Technology, and NEOWISE, which is a project of the Jet Propulsion Laboratory/California Institute of Technology. WISE and NEOWISE are funded by the National Aeronautics and Space Administration.\\

This research has made use of data obtained from the SuperCOSMOS Science Archive, prepared and hosted by the Wide Field Astronomy Unit, Institute for Astronomy, University of Edinburgh, which is funded by the UK Science and Technology Facilities Council.\\

Funding for SDSS-III has been provided by the Alfred P. Sloan Foundation, the Participating Institutions, the National Science Foundation, and the U.S. Department of Energy Office of Science. The SDSS-III web site is \url{http://www.sdss3.org/}. SDSS-III is managed by the Astrophysical Research Consortium for the Participating Institutions of the SDSS-III Collaboration including the University of Arizona, the Brazilian Participation Group, Brookhaven National Laboratory, Carnegie Mellon University, University of Florida, the French Participation Group, the German Participation Group, Harvard University, the Instituto de Astrofisica de Canarias, the Michigan State/Notre Dame/JINA Participation Group, Johns Hopkins University, Lawrence Berkeley National Laboratory, Max Planck Institute for Astrophysics, Max Planck Institute for Extraterrestrial Physics, New Mexico State University, New York University, Ohio State University, Pennsylvania State University, University of Portsmouth, Princeton University, the Spanish Participation Group, University of Tokyo, University of Utah, Vanderbilt University, University of Virginia, University of Washington, and Yale University.\\ 

MB, KM, AP, AK and MK were supported by the Polish National Science Center under contract \#UMO-2012/07/D/ST9/02785.
TK, KM and AP were supported by the National Science Centre (grants UMO-2012/07/B/ST9/04425 and UMO-2013/09/D/ST9/04030), the Polish-Swiss Astro Project (co-financed by a grant from Switzerland, through the Swiss Contribution to the enlarged European Union), and the European Associated Laboratory Astrophysics Poland-France HECOLS. 
MB was supported by the Netherlands Organization for Scientific Research, NWO, through grant number 614.001.451, and through FP7 grant number 279396 from the  European Research Council. 

\end{acknowledgements}

\bibliographystyle{aa}
\bibliography{TKrakowski}

\begin{appendix}

\section{Calibration of $W3$ and $W4$ upper limits}
\label{App: W3 calibration}

\begin{figure*}
\includegraphics[width=0.45\textwidth]{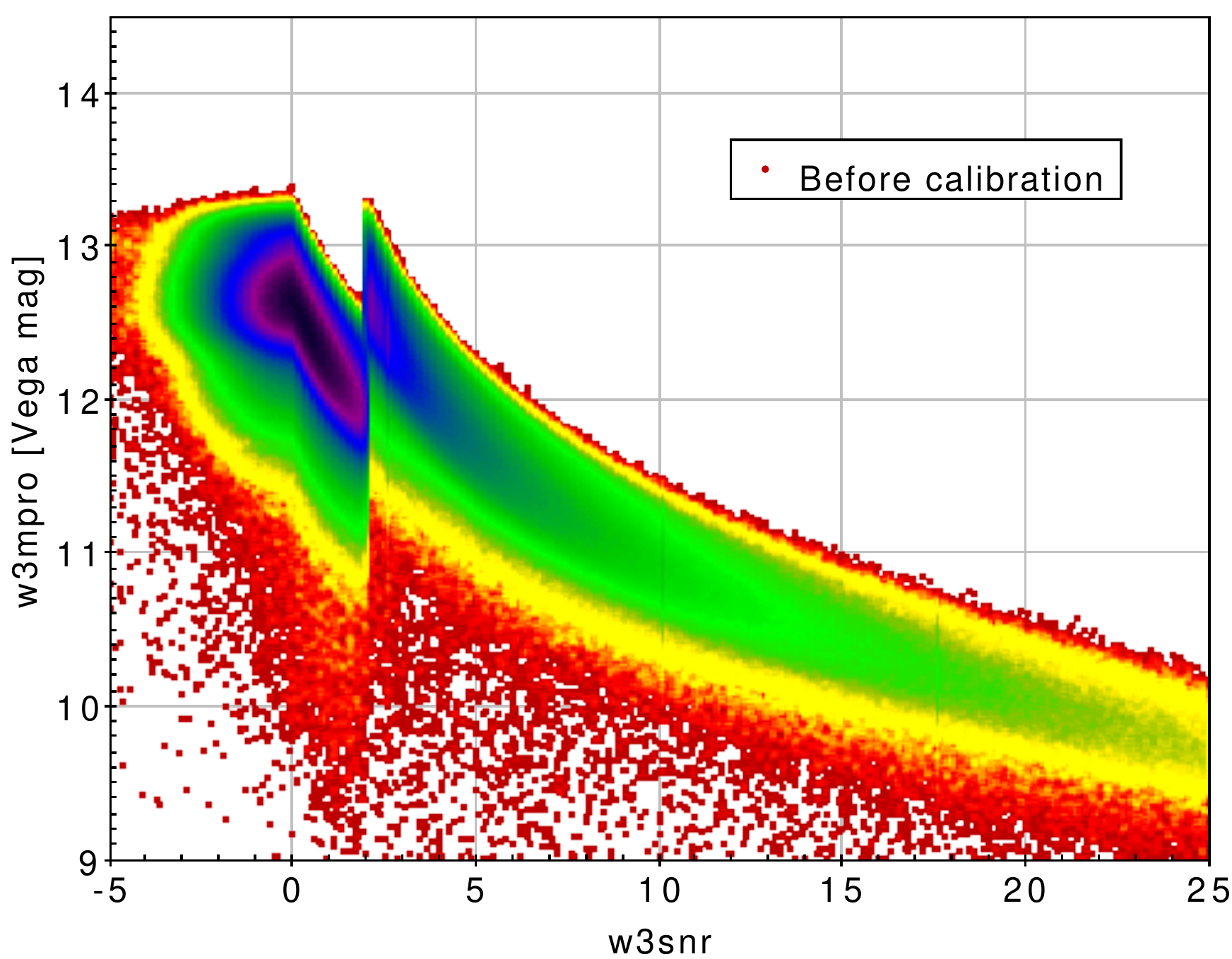} 
\includegraphics[width=0.45\textwidth]{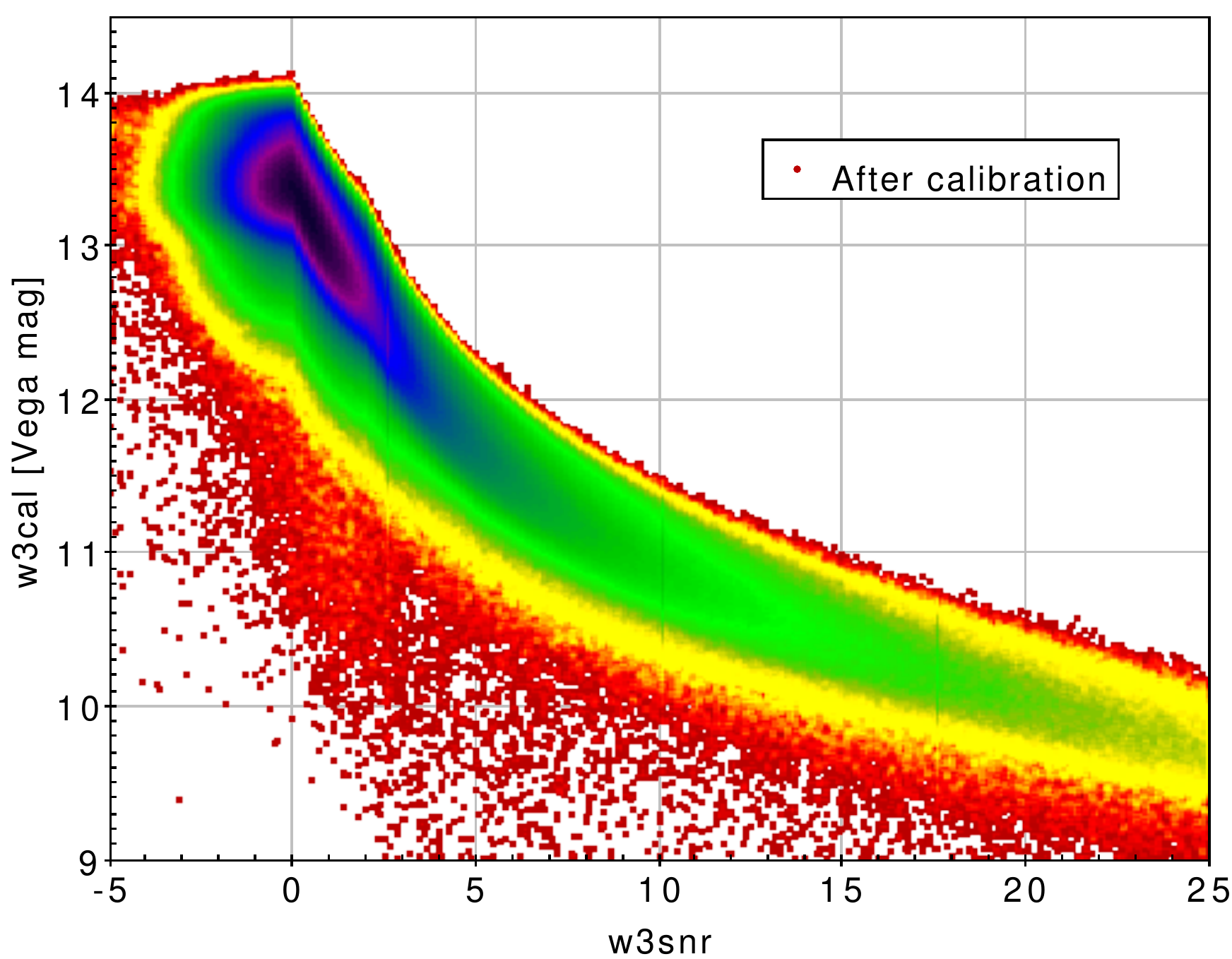} 
\caption{\label{Fig: W3 calibration}Illustration of the calibration procedure of $W3$ upper limits and non-detections: values from the database (left panel) and after our empirical offset  by $+0.75$ mag for the $\mathtt{w3snr}<2$ sources (right panel).}
\end{figure*}

\begin{figure*}
\includegraphics[width=0.45\textwidth]{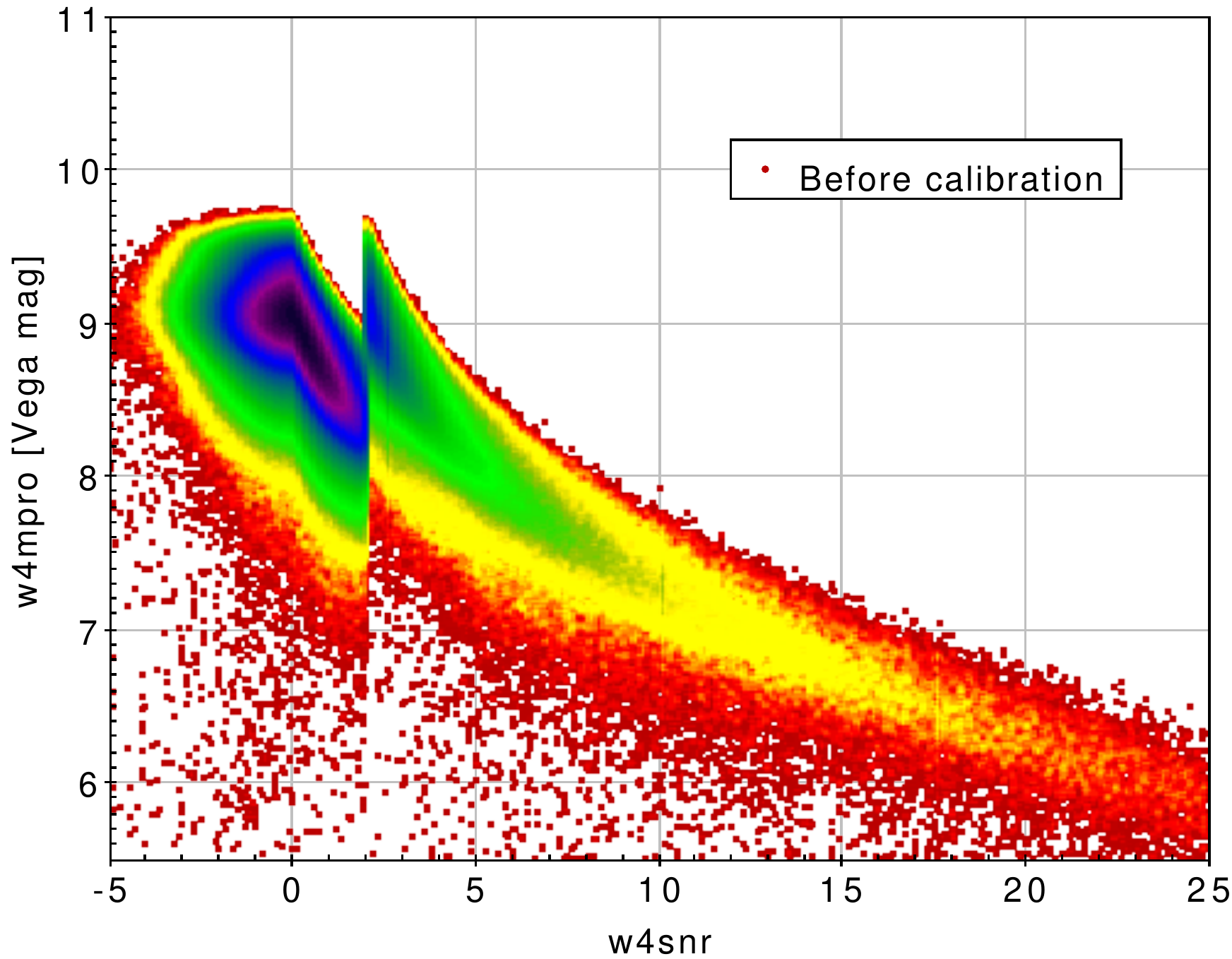} 
\includegraphics[width=0.45\textwidth]{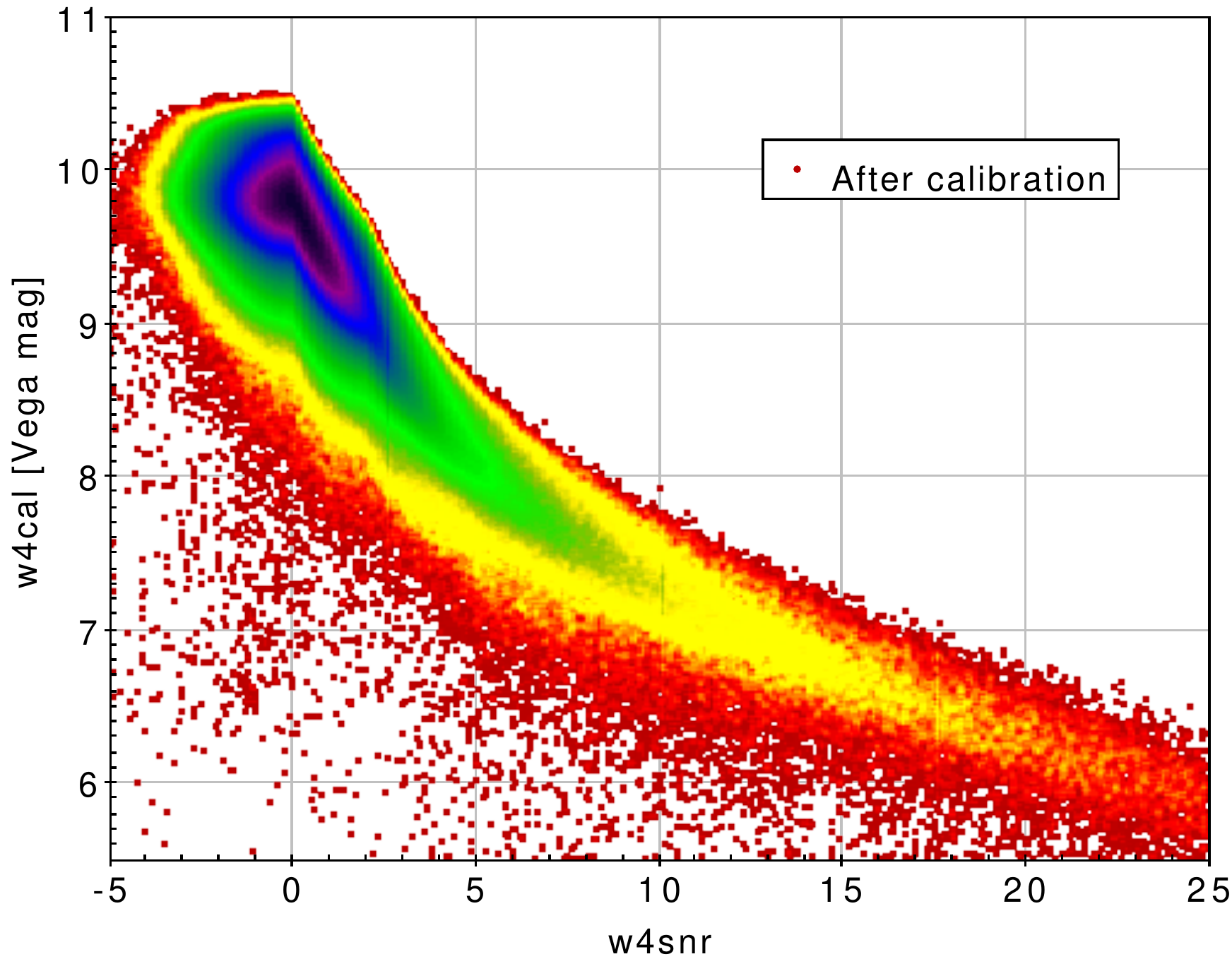} 
\caption{\label{Fig: W4 calibration}Illustration of the calibration procedure of $W4$ upper limits and non-detections: values from the database (left panel) and after our empirical offset by $+0.75$ mag for the $\mathtt{w4snr}<2$ sources (right panel).}
\end{figure*}

The AllWISE Source Catalogue lists only the sources that were detected with $S/N\geq 5$ in at least one of the four survey bands. On the other hand, whenever there was a $5\sigma$ detection in any of the bands, all the other bands were also measured at the given position, which means that each of the catalogued sources has magnitudes listed for all the bands (except for some very rare cases of processing or instrumental artefacts). Because of the much higher sensitivities in the $W1$ and $W2$ bands than in the two other channels, the AllWISE catalogue is mostly $W1$
selected. In the $W3$ channel (12 $\mu$m), most of the $\mathtt{w1snr}\geq 5$ sources will have $\mathtt{w3snr}<2$. Such objects are provided in the database as upper limits (or non-detections if $\mathtt{w3snr}<0$) and their fluxes are systematically overestimated. To be able to use such measurements in the classification procedure, we have designed an empirical correction for $W3$ upper limits and non-detections. Analysing the dependence of the mean $W3$ magnitude value on the $W3$ S/N estimate from the database, we have found that there is a roughly constant shift in \ttt{w3mpro} at the $\mathtt{w3snr}=2$ threshold, amounting to $\sim 0.75$ mag, which can therefore be removed by artificially dimming the low-\ttt{w3snr} sources. Figure \ref{Fig: W3 calibration} illustrate this calibration procedure for a random sample of WISE sources: the left panel shows quantities taken directly from the database, while the right panel presents our `\ttt{w3cal}' magnitude on the $y$-axis, obtained by adding $0.75$~mag to the database $\mathtt{w3mpro}$ value. Obviously, at $\mathtt{w3snr}<0$, the values are pure noise. A similar procedure can be applied to the $W4$ ($\sim 23$~$\mu$m) band, where the relevant offset for $\mathtt{w4snr}<2$ sources was found to be the same as in $W3$. This is illustrated in Fig. \ref{Fig: W4 calibration}. In this case, most of the sources remain undetected at all ($\mathtt{w4snr}<0$); for this reason, we did not use this band in our work.

We note that a more appropriate way of estimating magnitudes for the $W3$ and $W4$ upper limits and non-detections would be to use information from another band(s) in which a given source is detected with  $S/N>5$ (for instance from the SuperCOSMOS bands). This is the idea behind the aperture-matched photometry, employed by GAMA \citep{Wright16} and the forced-photometry technique applied to 400 million WISE sources selected from the SDSS \citep{LHS16}. This method, albeit certainly of great interest, is beyond the scope of the present work, however.

\end{appendix}

\end{document}